\documentclass[table,twocolumn]{aastex62}
\usepackage{amstext}
\usepackage{amsmath}
\usepackage{apjfonts}
\usepackage{graphicx}
\usepackage[capbesideposition={top}]{floatrow}
\usepackage{float}
\usepackage{cancel}
\usepackage{floatrow}
\usepackage{cleveref}
\usepackage{booktabs}
\usepackage{hyperref}

\newcommand{\beqar}{\begin{eqnarray}}
\newcommand{\eeqar}{\end{eqnarray}}

\newcommand{\beq}{\begin{equation}}
\newcommand{\eeq}{\end{equation}}

\definecolor{nick}{HTML}{006400}
\definecolor{callum}{HTML}{da144e}

\begin{document}
\title{Little Red Dots As Super-Eddington Fountain Flows}

\correspondingauthor{Nicholas Kaaz}
\email{nkaaz@princeton.edu}

\author[0000-0002-5375-8232]{Nicholas Kaaz}
\affiliation{Princeton Center for Theoretical Sciences, Princeton University, Princeton, NJ 08544, USA}
\affiliation{Princeton Gravity Initiative, Princeton University, Princeton, NJ 08544, USA}

\author[0000-0001-9185-5044]{Eliot Quataert}
\affiliation{Department of Astrophysical Sciences, Princeton University, Peyton Hall, Princeton, NJ 08540, USA}

\author{Taya Govreen-Segal}
\affiliation{School of Physics and Astronomy, Tel Aviv University, Tel Aviv 6997801, Israel}
\affiliation{Department of Astrophysical Sciences, Princeton University, Peyton Hall, Princeton, NJ 08540, USA}

\author[0000-0003-2488-4667]{Hanpu Liu}
\affiliation{Department of Astrophysical Sciences, Princeton University, Peyton Hall, Princeton, NJ 08540, USA}

%\author[0000-0001-9185-5044]{Eliot Quataert}
%\affil{Department of Astrophysical Sciences, Princeton University, Princeton, NJ 08544, USA} 

\begin{abstract}
Little red dots (LRDs) may be powered by supermassive black holes (SMBHs) accreting above the Eddington limit. Spectroscopy of LRDs often shows absorption troughs blueshifted by $\sim100-300\,{\rm km\,s^{-1}}$, implying a slow wind. This is puzzling: super-Eddington disks drive much faster winds, $\gtrsim10^3-10^4\,{\rm km\,s^{-1}}$. We argue that LRDs are super-Eddington SMBHs viewed off-axis and engulfed in a slow wind that covers most sight-lines; the fast wind escapes near the poles. Trapped light puffs the inner disk into a quasi-spherical envelope that launches the slow wind. The wind may be ``photon-tired'', meaning the light only barely unbinds it. Such marginally unbound winds lead to fountain flows, where some gas escapes and the rest falls back. We model the envelope with idealized, spherically symmetric ``marginally unbound'' ($v\sim v_{\rm esc}$) and ``photon-tired'' winds. We feed these profiles into the radiative transfer code \textsc{Sirocco} to study how the wind reprocesses the light from an accreting $10^6\,M_\odot$ SMBH. We describe most of the wind as a ``Balmer cocoon'' -- a Compton-thick region in which depletion of Balmer continuum photons ($h\nu >3.4\,{\rm eV}$) rather than Lyman continuum photons ($h\nu > 13.6\,{\rm eV}$) keeps the gas ionized. The spectra span the range of LRD-like sources: ``little blue dots'' at lower outflow rates ($\sim2.5\,M_\odot\,{\rm yr}^{-1}$); V-shaped LRDs with a Balmer break ($\sim5-10\,M_\odot\,{\rm yr}^{-1}$); and red LRDs with full breaks ($\sim15\,M_\odot\,{\rm yr}^{-1}$). Our Balmer line profiles show P~Cygni features atop broad, exponential wings, as is observed. The break is possible at lower wind densities than in LTE models (or NLTE models without electron scattering). This is because Lyman~$\alpha$ trapping sustains our $n=2$ hydrogen population and electron scattering enhances the optical depth. Our arguments are also applicable to supermassive stars or quasi-stars, which may launch winds with similar properties.  
\end{abstract}

%%%%%%%%%%%%%%%%%%%%%%%%%%%%%%
%% SECTION 1 : INTRODUCTION %% 
%%%%%%%%%%%%%%%%%%%%%%%%%%%%%%

\section{Introduction}
\label{sec:intro}
``Little red dots'' (LRDs) \citep{matthee_2024} are enigmatic sources that are abundant in the early Universe \citep{furtak_2023,labbe_2023,kokorev_2024,kocevski_2025,labbe_2025}. LRDs are bright and compact, suggesting that they may be active galactic nuclei (AGN). Yet, LRDs do not look like typical, UV-bright AGN; they lack detectable X-ray \citep{ananna_2024,yue_2024,sacchi_2025} and radio emission \citep{akins_2025,perger_2025,mazzolari_2026}, show little to no variability, and appear red rather than blue in the rest-frame. Although these dissimilarities have motivated alternative progenitors, AGN remain viable if there is a great deal of gas that envelops the accreting supermassive black hole (SMBH). This hints that LRDs may be exotic AGN that harbor SMBHs accreting at high rates, a possibility we explore in this work. 

%% FIGURE CARTOON
\begin{figure*}[bht]
    \centering
    \includegraphics[width=\textwidth]{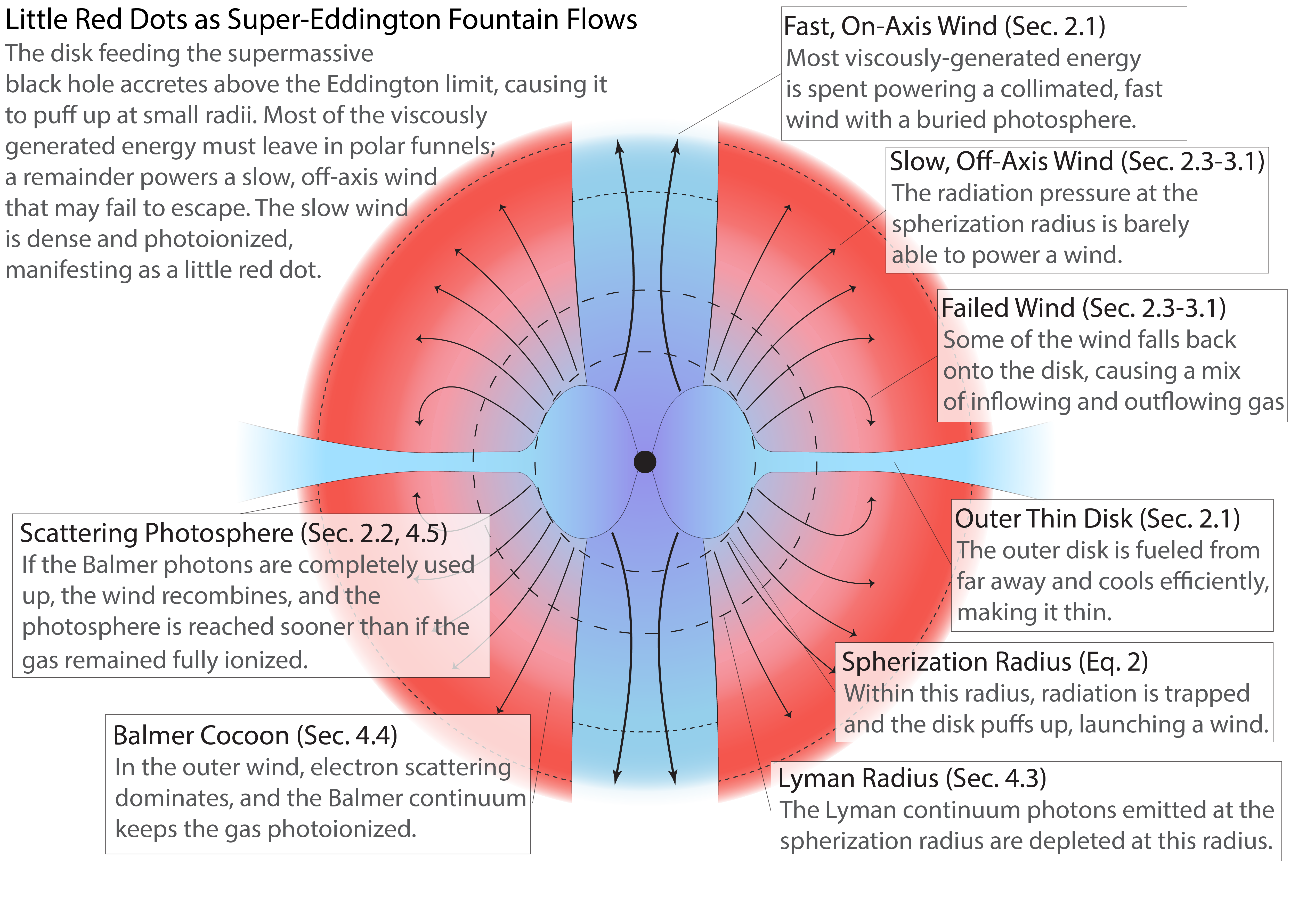}
    \caption{Infographic describing the key features of our model.}
    \label{fig:cartoon}
\end{figure*}

Why are LRDs red? One possibility is that the AGN is heavily enshrouded by dust. In UV-bright AGN, some of the blue light from the inner disk is reprocessed by a dusty torus that re-emits at $500-2000\,{\rm K}$ \citep{edelson_1986,barvainis_1987,sanders_1989}. This emission is not detected in LRDs \citep{casey_2024,williams_2024,setton_2025,casey_2025,degraaff_2025}, suggesting that either dust is present but emits in narrowly-missed spectral ranges \citep[e.g.,][]{setton_2025} or that LRDs are intrinsically red \citep[although, see][]{madau_maiolino_2026}. If one abandons dust, the inferred bolometric luminosities are near the Eddington limit for a $10^6\,M_\odot$ SMBH \citep{greene_2026}, which is lower than the large SMBH masses ($\lesssim10^8\,M_\odot$) required in the dusty AGN scenario \citep{lin_2024,matthee_2025}. Lower SMBH masses are easier to reconcile with existing SMBH growth and AGN feedback models \citep{luberto_furlanetto_2025} because lighter SMBHs are easier to assemble when the Universe is young. 

If the redness of LRDs is intrinsic, one clue to their origins is that their red continua fall off near the Balmer break (sometimes gradually, sometimes sharply). This only occurs if enough hydrogen in the $n=2$ state reprocesses the light emitted by the central engine. For instance, \cite{inayoshi_2025} found that a Balmer break emerges when a thick slab of $\gtrsim 10^{9}\,{\rm cm^{-3}}$ gas attenuates a standard quasar spectral energy distribution (SED). Here, they attribute the high density of the $n=2$ state to collisions. \cite{liu_2025} imagined SMBHs enshrouded by large spheres of gas in local thermodynamic equilibrium \citep[see also][]{liu_2026}. These structures have surface temperatures near $4000-7000\,{\rm K}$ and exhibit Balmer breaks when the photospheric density is $\lesssim10^{13}\,{\rm cm^{-3}}$. The surface temperature depends only weakly on density because the opacity drops sharply below $10^4$ K, acting as a thermostat. This may be why LRDs are ubiquitously cool (stars near the Hayashi limit behave similarly).

LRDs also have blue emission. The blue and red components together produce a characteristic V-shape, where the minimum is near the Balmer break \citep{setton_2025_vshape}. Both the AGN and the host galaxy may contribute to the blue emission. Some have advocated that the enshrouded AGN is \textit{completely} red \citep[``BH stars'',][]{naidu_2025,degraaff_2025_bhs, sun_2026} and that the blue emission is solely from the galaxy. The BH star hypothesis is supported by individual LRDs that show extremely faint blue emission \citep{degraaff_2025,naidu_2025}. However, other LRDs show high-ionization lines \citep{tang_2025,tripodi_2025b,ji_2026}, which are typical of UV-bright AGN. It may then be that some LRDs have intrinsic blue emission that escapes and some do not; a complete picture of LRDs should accommodate both. 

LRD spectra feature broad ($\gtrsim3000\,{\rm km\,s^{-1}}$) emission lines \citep{harikane_2023,kocevski_2023,matthee_2024}. In canonical AGN, the broad lines are produced by gas photoionized by the blue continuum. The lines are broad because the line-emitting gas is rotating (or outflowing). However, the wings of the broad lines in LRDs are better fit by exponentials \citep{rusakov_2026,kokorev_2026,matthee_2026} (broad lines in UV-bright AGN are usually, but not always, well-described by Gaussians or Lorentzians). When emission-line photons repeatedly scatter, they develop (approximately) exponential wings even if the gas is static \citep{laor_2006,Elisha2026}.\footnote{\citet{Elisha2026} show that the wings are exponential only if the scattering medium is external to the line emitting region; when the two are co-spatial the wings are broken power-laws.   We defer trying to distinguish between these two line profiles in our models (or the data) to future work and refer to the line wings as `exponential' throughout this paper.}  This is especially important in highly super-Eddington AGN, which launch dense outflows via radiation pressure on electrons. Then, the width of the emission lines produced within these outflows is not necessarily set by the velocity of the line-emitting gas \citep[although, see][]{scholtz_2026,madau_2026,brazzini_2026}. Still, LRD emission lines have narrower features that may be set by the kinematics of the gas. For instance, the emission lines feature narrow (``P~Cygni'') absorption troughs that are typically blueshifted \citep[or, more rarely, redshifted, e.g.][]{matthee_2024} by $\lesssim300\,{\rm km\,s^{-1}}$ from line center \citep{juodzbalis_2024,lin_2024,degraaff_2025,deugenio_2025,kocevski_2025,naidu_2025,wang_2025,deugenio_2026a,deugenio_2026b,kokorev_2026,yanagisawa_2026}. These velocities are hard to explain, as they are much slower than those measured in UV-bright AGN; this puzzle motivates much of what follows. 

%% FIGURE OUTFLOW PROBLEM
\begin{figure*}[bht]
    \centering
    \includegraphics[width=0.9\textwidth]{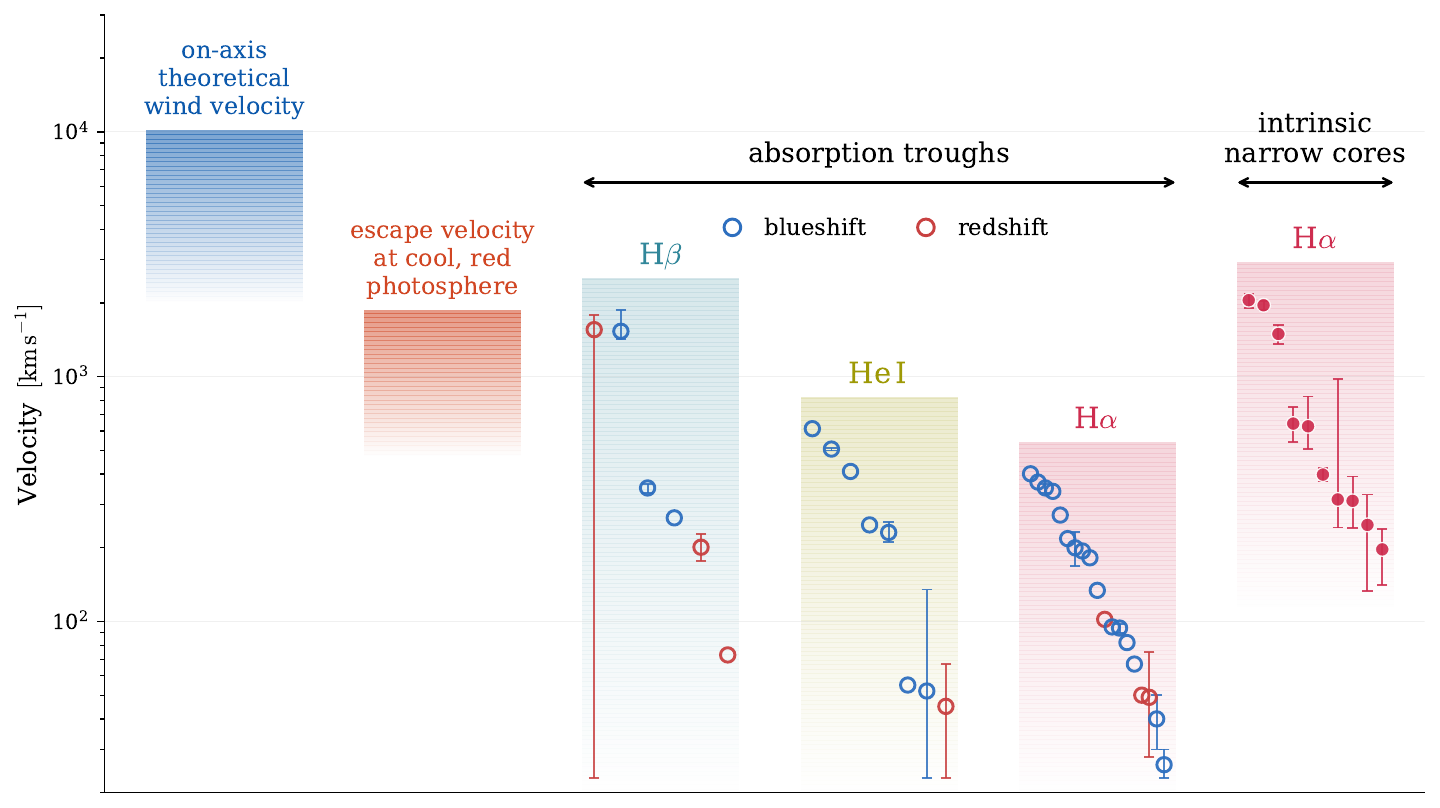}
    \caption{Summary of velocities inferred from observations and theoretical expectations. The discrepancy between the two suggests that LRD winds are slow and marginally unbound (Sec.~\ref{sec:fountain_flow}), possibly due to photon tiring (Sec.~\ref{sec:photon_tired_winds}). We list the theoretical velocity of the super-Eddington winds launched within $r<r_{\rm sph}$ (Eq.~\ref{eq:velocity_onaxis}), which we suggest are collimated and launched on-axis (see the funnels drawn in Fig.~\ref{fig:cartoon}). We also list the escape velocity from a thermal photosphere given typical LRD color temperatures $T\sim3000-7000\,{\rm K}$ and $L=L_{\rm Edd}$. Both of these estimates exceed the slow velocities inferred from the observed line profiles. We list the offset velocity of blueshifted (and sometimes redshifted) absorption troughs for H$\alpha$, H$\beta$ and He I (Table~\ref{tab:lrd_absorption} in Appendix~\ref{app:obs_vel}). We also list the FWHM velocities associated with the width of the intrinsic narrow H$\alpha$ cores from \cite{rusakov_2026}.}
    \label{fig:velocities}
\end{figure*}

The purpose of this paper is to provide a qualitative physical picture, supported by radiative transfer calculations, of LRDs. We propose that they are SMBHs accreting well above the Eddington limit, viewed off-axis, with slow, marginally unbound outflows that barely escape an inflated photosphere; see Fig.~\ref{fig:cartoon} for an infographic. Although we focus on the SMBH model, many of the basic arguments and calculations we present would also apply to similar winds driven by a supermassive star or a quasi-star.    In Sec.~\ref{sec:outflow_problem}, we motivate the need for a marginally unbound wind. In Sec.~\ref{sec:lrds_and_stars}, we draw analogies between LRDs and photon-tired winds from Eddington-limited massive stars. In Sec.~\ref{sec:radtransfer}, we describe radiative transfer calculations of marginally unbound and photon-tired wind models. In Sec.~\ref{sec:disc} we discuss our results, mention a few caveats, and summarize.

%%%%%%%%%%%%%%%%%%%%%%%%%%%%%%%%%%%%%
%% SECTION 2 : The Outflow Problem %% 
%%%%%%%%%%%%%%%%%%%%%%%%%%%%%%%%%%%%%

\section{The outflow problem}
\label{sec:outflow_problem}

The broad ($\gtrsim 3000\,{\rm km\,s^{-1}}$), exponential wings of LRD emission lines \citep{rusakov_2026} are likely produced by repeated scatterings of photons \citep{laor_2006}, rather than by Doppler shifts due to the velocity of the line-emitting gas. This is reasonable, because if LRDs are SMBHs accreting well above the Eddington limit, then they should launch outflows via radiation pressure on electrons. These scattering-thick outflows may produce emission lines. 

If LRD emission lines are broadened by scattering, is there another way to probe the velocity of the line-emitting gas? LRD emission lines often feature narrow (``P~Cygni'') absorption troughs that are blueshifted from line center by $\lesssim300\,{\rm km\,s^{-1}}$ \citep{matthee_2026}. These troughs occur when an absorbing shell is backlit by a broadened emission line, and the blueshift tells us how fast the shell is moving. In Fig.~\ref{fig:velocities}, we collect the Doppler shifts of absorption troughs from various sources (see Table~\ref{tab:lrd_absorption} in the Appendix). We also include the narrow-line core widths from the models in \cite{rusakov_2026}, which they determined by assuming that the exponential wings result from the scattering of an intrinsic narrow core. Interestingly, the absorption troughs are sometimes redshifted rather than blueshifted, indicating that the absorbing gas is moving inwards. Furthermore, the broad scattering wings are symmetric. This also constrains the velocity of the line-emitting gas, since, if it were purely outflowing (or inflowing), the scattering wings would be asymmetric. Thus, the line-emitting gas must have both inflowing and outflowing components \citep{sneppen_2026a}.

\subsection{Fast AGN winds}
\label{sec:fast_agn_winds}

The slow velocities in LRDs are surprising in the context of black hole accretion models. We usually infer much larger, $\gtrsim10^3-10^4\,{\rm km\,s^{-1}}$ velocities in AGN \citep{king_pounds_2015}. This is a straightforward consequence of the strong gravitational potential because outflow velocities are typically near the local escape velocity at their footprint. Consider a $10^6\,M_\odot$ SMBH that is accreting from a disk that extends to large radii. The disk cools efficiently in a region\footnote{At larger radii, the disk becomes unstable to star formation, but this is not important near $r_{\rm sph}.$} that is not too close (and not too far) from the horizon. In this region, the disk is thin. The local viscous dissipation rate per unit area is \citep{pringle_1981},
\begin{equation}
    D(r)=\frac{3}{8\pi}\dot{M}\Omega^2,
\end{equation}
where $\Omega$ is the orbital frequency (assumed to be Keplerian) and $\dot{M}$ is the mass accretion rate.
The emitted light exerts a vertical force on the electrons, $F_{\rm rad,z}(r)\approx\kappa_{\rm es}D(r)/c$ (where $\kappa_{\rm es}=0.34\,{\rm cm^2}\,{\rm g}^{-1}$). At small radii, this force competes with gravity, which exerts force $F_{\rm grav,z}\approx \Omega^2H$. Here, $H$ is the disk scale height. These two forces are equal at the ``spherization radius'',
\begin{equation}
    r_{\rm sph}\approx 1.4\times10^{15}\,{\rm cm}\left(\frac{\dot{M}}{5\,M_\odot\,{\rm yr}^{-1}}\right)\left(\frac{h}{0.3}\right)^{-1},
    \label{eq:spherization_radius}
\end{equation}
where $h=H/r$. Within $r_{\rm sph}$, the disk becomes quasi-spherical and radiatively inefficient \citep{ss73}. The transition from thin to thick at $r_{\rm sph}$ is illustrated in Fig.~\ref{fig:cartoon}. The fiducial accretion rate of $5\,M_\odot\,{\rm yr}^{-1}$ is highly super-Eddington,
\begin{equation}
    \dot{M}\approx 190 \dot{M}_{\rm Edd},
\end{equation}
where
\begin{equation}
    \dot{M}_{\rm Edd}\approx 0.026\,M_\odot\,{\rm yr}^{-1}\left(\frac{M}{10^6\,M_\odot}\right)\left(\frac{\eta}{0.1}\right)^{-1}
\end{equation}
is the Eddington accretion rate and $\eta$ is a fiducial radiative efficiency used to define $\dot M_{\rm Edd}$ (not necessarily the true radiative efficiency of the flow). At this high accretion rate, $r_{\rm sph}$ is about five thousand Schwarzschild radii away from the horizon. Within $r_{\rm sph}$, the viscously-generated energy that drives accretion is lost in a wind. The wind robs the disk of matter so that the accretion rate obeys a power law with radius,
\begin{equation}
    \dot{M}_{\rm in}(r) \approx \dot{M}_{\rm BH} + \dot{M}(r/r_{\rm sph})^p,\qquad (r\leq r_{\rm sph})
\end{equation}
where $\dot{M}_{\rm BH}\ll \dot{M}$ is the leftover mass that makes it to the horizon. Most of the mass is deposited in outflows,
\begin{equation}
    \dot{M}_{\rm out}(r) \approx \dot{M}(r/r_{\rm sph})^p,\qquad (r\leq r_{\rm sph})
\end{equation}
unless $p$ is unrealistically small. At every radius, the wind launches at a velocity that cannot greatly exceed the local escape velocity,
\begin{equation}
    d\dot{E}_{\rm out}\approx\frac{GM}{r} d\dot{M}_{\rm out}\approx p\frac{GM}{r_{\rm sph}}\dot{M}  (r/r_{\rm sph})^{p-1} r^{-1}dr
\end{equation}
If $p<1$, most of the mass leaves the disk near $r_{\rm sph}$, but most of the energy leaves at small radii. The exponent $p=1$ is special, because then the amount of energy released at every decade in radius is equal. Values $p>1$ are thought to be unphysical \citep{adios99}; indeed, simulations of radiatively inefficient accretion flows ubiquitously show that $\frac{1}{2}\lesssim p\lesssim1$ \citep{pang_2011,macleod_2015,ressler_2018,xu_2019,koushik_2022,xu_2023,vikram_2024,minghao_2025,hyerin_2025,aris_2025}. 

How fast are these outflows? First, imagine that the outflows do not mix, and that the terminal velocity is virial at the launch point. Then, the slowest streamlines still escape at no less than $\sqrt{2GM/r_{\rm sph}}\sim 4300\,{\rm km\,s^{-1}}$ for our fiducial parameters in Eq.~\ref{eq:spherization_radius}, while the fast on-axis streamlines may be mildly relativistic. Alternatively, if we take $p=1$ and assume that the winds are well-mixed (so that slower, heavier outflows from large radii weigh down faster, lighter outflows from small radii), then the entire outflow expands at $v_{\rm out}^2\approx \int \mathrm{d}\dot{E}_{\rm out}/\int \mathrm{d}\dot{M}_{\rm out}$. This expression reduces to roughly twice the escape velocity at $r_{\rm sph}$, 
\begin{equation}
    v_{\rm out} \gtrsim 9200\,{\rm km\,s^{-1}} \left(\frac{r_{\rm sph}}{1.6\times10^{15}\,{\rm cm}}\right)^{-1/2}\left(\frac{M}{10^6\,M_\odot}\right)^{1/2},
    \label{eq:velocity_onaxis}
\end{equation}
which is also far in excess of LRD wind velocities (see Fig.~\ref{fig:velocities}). This issue is only circumvented if the accretion rate balloons to $\gtrsim10^3-10^4\,M_\odot\,{\rm yr}^{-1}$ -- then, $r_{\rm sph}$ would be much larger, and the escape velocity would be much lower. However, such large rates greatly exceed the estimated accretion rates onto the lower-mass dark matter halos \citep{claudeandre_2011,dekele_2017} that host LRDs \citep{arita_2025,pizzati_2025,lin_2026}. 

\subsection{Extended photospheres of LRDs}
\label{sec:lrd_photosphere}

Forget, for a moment, about the fast outflows, and imagine the LRD as an accreting SMBH buried in a dense, spherical envelope of unknown origin. The cool, red temperatures imply a thermal photosphere radius, 
\begin{equation}
    r_{\rm th}\approx 10^{16}\,{\rm cm} \left(\frac{T_{\rm eff}}{6100\,{\rm K}}\right)^{-2}\left(\frac{L}{10^{44}\,{\rm ergs\,s^{-1}}}\right)^{1/2},
    \label{eq:thermal_photosphere}
\end{equation}
where the escape velocity is $\approx 1600\,{\rm km\,s^{-1}}$. This is still well above LRD velocities. However, the scattering photosphere must be much larger than the thermal photosphere because the electron scattering opacity is much greater than the absorption opacity at the densities we care about. Given Kramers bound-free opacity at solar metallicity and an example density profile $\rho\propto r^{-2}$, we find
\begin{equation}
\begin{aligned}
    r_{\rm sca}&\approx r_{\rm th} \sqrt{\frac{\kappa_{\rm sca}}{3\kappa_{\rm abs}}}\\&\approx 10^2r_{\rm th} \left(\frac{\rho(r_{\rm th})}{10^{-14}\,{\rm g\,cm^{-3}}}\right)^{-1/2}\left(\frac{T_{\rm e}}{10^4\,{\rm K}}\right)^{7/4},
\end{aligned}
\label{eq:r_sca_inferred}
\end{equation}
where $T_{\rm e}$ is the electron temperature, which we assume is fixed beyond $r_{\rm th}$, and $\rho$ is evaluated at $r_{\rm th}$. The extended scattering photosphere is only possible because the density is so low; in a stellar envelope,
$\kappa_{\rm abs}>\kappa_{\rm sca}$ at $10^4\,{\rm K}$. This is attractive, as it naturally provides a region where emission lines can both form and scatter repeatedly, developing exponential wings in their observed profiles. Furthermore, at $r_{\rm sca}$, the escape velocity is $\approx 160\,{\rm km\,s^{-1}}$. This is suggestively in line with the observed velocities; the absorption troughs may form just beyond the scattering photosphere. However, the gas may instead recombine before the scattering photosphere is reached, which we find in Sec.~\ref{sec:radtransfer}. Later, it will be clear that the grey Kramers opacity is a poor approximation in LRDs because the opacity's frequency-dependence is essential. Regardless, the scattering opacity still dominates.

How can the AGN have a slowly outflowing, extended, scattering-dominated envelope if a fast wind must launch from $r<r_{\rm sph}$? There are two resolutions. The first is viewing angle: much of the energy generated within $r_{\rm sph}$ should escape as a wind, so if we observe a slow wind, then the fast wind must escape near the poles. Viewed on-axis, the AGN might appear quasar-like \citep[for other viewing angle arguments, see][]{madau_maiolino_2026,zhou_2026}. The second resolution is the acceleration of the wind. It must either occur near the scattering photosphere, or, if the acceleration occurs deep within the photosphere, then it must be inefficient, such that the gas barely escapes. 

\subsection{Marginally unbound flows}
\label{sec:fountain_flow}

\begin{figure}%[bht]
    \centering
    \includegraphics[width=\textwidth]{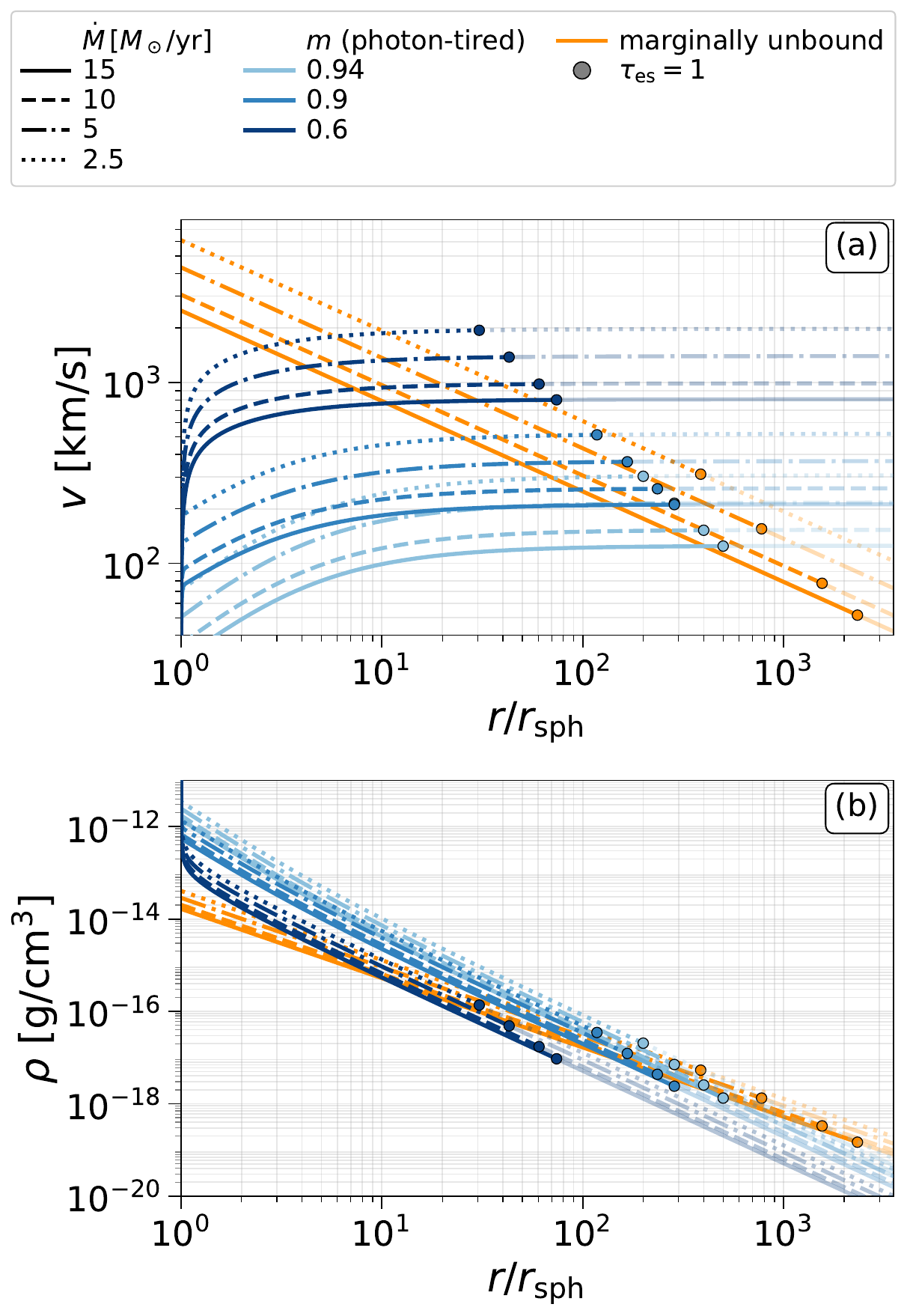}
    \caption{Velocity and density profiles for marginally unbound (Sec.~\ref{sec:fountain_flow}) and photon-tired (Sec.~\ref{sec:photon_tired_winds}) winds. We consider outflow rates $\dot{M}_{\rm out}=2.5-15\,M_\odot\,{\rm yr}^{-1}$ in both, and for the photon-tired winds we consider tiring parameters ($m$, Eq.~\ref{eq:tiring_parameter_redd}) that are near unity (at $m>1$, the wind fails). In either wind, the (fully-ionized) scattering photosphere (scatter points) is roughly at $\sim10^2-10^3\,r_{\rm sph}$ ($\sim 10^{17}-10^{18}\,{\rm cm}$), assuming fully ionized gas; in practice, the gas may recombine, keeping the scattering photosphere somewhat smaller than estimated here, especially at higher densities. As $m$ decreases, the velocity becomes too fast to explain LRDs.}
    
    \label{fig:wind_profiles}
\end{figure}

Imagine a ``marginally unbound'' wind, launched from $r_{\rm sph}$, that barely escapes the gravitational potential and has outflow rate $\dot{M}_{\rm out}$. Here, we neglect the fast wind and assume spherical symmetry. The entire marginally unbound wind propagates exactly at the escape velocity,
\begin{equation}
    v^{\rm (mu)}(r)\approx 163\,{\rm km\,s^{-1}}\left(\frac{M}{10^6\,M_\odot}\right)^{1/2}\left(\frac{r}{10^{18}\,{\rm cm}}\right)^{-1/2},
    \label{eq:v_mu}
\end{equation}
and the density profile is fixed by mass conservation to be,
\begin{equation}
\begin{aligned}
    \rho^{\rm (mu)}(r) &\approx 1.6\times 10^{-18}\,{\rm g\,cm^{-3}}\\\times&\left(\frac{\dot{M}_{\rm out}}{5\,M_\odot\,{\rm yr}^{-1}}\right)\left(\frac{M}{10^6\,M_\odot}\right)^{-1/2}\left(\frac{r}{10^{18}\,{\rm cm}}\right)^{-3/2}
\end{aligned}
\end{equation}
The scattering photosphere of the marginally unbound (and fully ionized) wind is at radius
\begin{equation}
r_{\rm sca}^{\rm (mu)}\approx 10^{18}\,{\rm cm}\left(\frac{\dot{M}_{\rm out}}{5\,M_\odot\,{\rm yr}^{-1}}\right)^{2}\left(\frac{M}{10^6\,M_\odot}\right)^{-1}.
\end{equation}
Encouragingly, both $r_{\rm sca}^{\rm (mu)}$ and the escape velocity at this radius,
\begin{equation}
    v^{\rm (mu)}(r_{\rm sca}^{\rm (mu)})\approx 156\,{\rm km\,s^{-1}}\left(\frac{\dot{M}_{\rm out}}{5\,M_\odot\,{\rm yr^{-1}}}\right)^{-1}\left(\frac{M}{10^6\,M_\odot}\right),
\end{equation}
are comparable to the size and velocity of the scattering photosphere inferred in Sec.~\ref{sec:lrd_photosphere}. We show the density and velocity profiles for a few values of $\dot{M}_{\rm out}$ in Fig.~\ref{fig:wind_profiles} (here, we also plot the ``photon-tired'' wind profiles that we describe in Sec.~\ref{sec:photon_tired_winds}). 

If the wind is truly ``marginally unbound'', then it will struggle to escape. Streamlines that are precisely marginally unbound have zero energy; in reality, each streamline should vary in energy, such that they obey a distribution with a mean near zero. Then, some streamlines escape, and some fail and fall back -- forming a ``fountain flow'' (illustrated in Fig.~\ref{fig:cartoon}). Fountain flows are especially appealing because they satisfy empirical constraints on LRD kinematics. The absorption troughs appear either blueshifted or redshifted depending on whether our line of sight intersects with inflowing or outflowing gas. Furthermore, some of the escaping photons will be scattered by inflowing gas, and some will be scattered by outflowing gas; this symmetrizes the scattering wings of the emission line profiles \citep[e.g.,][]{sneppen_2026a}. 

These arguments suggest that a marginally unbound fountain flow launched from $r_{\rm sph}$ is phenomenologically consistent with LRD envelopes. We will later find that their SEDs and H$\alpha$ line profiles match, as well. Still, this structure might appear dynamically ad hoc. Why would the fountain flow emerge at all? 

%% FIGURE WIND PROFILES

%%%%%%%%%%%%%%%%%%%%%%%
%% SECTION 2 : MASSIVE STAR ANALOGY %% 
%%%%%%%%%%%%%%%%%%%%%%%

\section{The Massive Star Analogy}
\label{sec:lrds_and_stars}

To glean insights into LRD envelopes, we borrow intuition from massive stars radiating near the Eddington limit. In the dense inner regions, energy is transported by convection. In the dilute outer regions, convection is inefficient, so the radiation field must diffuse outwards or launch a wind. The ``effective'' gravitational force is,
\begin{equation}
    g_{\rm eff} = -(1-\Gamma)\frac{GM}{r^2}, 
\end{equation}
where
\begin{equation}
    \Gamma = \frac{L}{L_{\rm Edd}}
\end{equation}
is the Eddington ratio. If $\Gamma$ is near unity, the effective gravity becomes small. The envelope is then weakly bound and may inflate  \citep[as in Wolf-Rayet stars and luminous blue variables; see][]{ishii_1999,petrovic_2006,grafener_2012}. If $\Gamma\lesssim 1$, the envelope is hydrostatic. If $\Gamma\gtrsim 1$, a wind is driven; the wind is slow and heavy if $\Gamma$ is not too large. Often, $\Gamma$ will transition past unity somewhere in the envelope \citep[due to a change in opacity,\footnote{The change in opacity may be genuine, usually due to the iron (or Helium) opacity bump at certain temperatures. There may instead be a change in the ``effective'' opacity due to porosity \citep{shaviv_2001}, which reduces the opacity locally if the medium is clumpy.} e.g. ][]{ro_matzner_2016}, and the solution changes from (nearly) hydrostatic to supersonic. However, this is difficult to achieve in LRDs where the scattering opacity is far greater than the absorption opacity (Eq.~\ref{eq:r_sca_inferred}), washing out any opacity bumps.  

In a super-Eddington accretion disk, the flow at $r<r_{\rm sph}$ is quasi-spherical and vigorously convective \citep{eliot_2000} -- much like the convective core of a massive star. The ``core'' of the disk is pressurized and wants to expand. Convection alone will transport material outwards where it may be accelerated by strong radiation forces. In analogy with massive, super-Eddington stars, we  propose that radiation pressure inflates an envelope and drives a wind. The inflated envelope will collimate the polar outflows and develop a large covering factor. In this picture, we would observe the LRD from most lines of sight.

\subsection{Photon-tired Winds}
\label{sec:photon_tired_winds}

When radiation pressure launches a wind, the kinetic energy of the wind comes at the expense of the radiation energy. Usually, this effect negligibly perturbs the radiation budget. However, when the mass loss rate is high, the radiative luminosity might barely be large enough to unbind the outflowing gas \citep{owocki_1997}. Then, the flow expends an order-unity fraction of the luminosity to unbind the wind. This is a statement of energy conservation and, for a wind launched at $r_{\rm sph}$, can be expressed as
\begin{equation}
    L(r) = L_0-\dot{M}_{\rm out}\left[\frac{v^2}{2}-\frac{GM}{r}+\frac{GM}{r_{\rm sph}}+\frac{4P_{\rm rad}}{\rho}\right].
    \label{eq:energy_conservation}
\end{equation}
Here, we neglected gas pressure because it is subdominant to radiation pressure (it is implicit that the sonic point is at some radius $r<r_{\rm sph}$). The radiation field will accelerate the wind as long as $L(r)>L_{\rm Edd}$. There is a characteristic $\dot{M}_{\rm out}$ at which the excess radiation budget barely unbinds the flow,
\begin{equation}
    \dot{M}_{\rm tir}\equiv (L_0-L_{\rm Edd})\frac{r_{\rm sph}}{GM}.
    \label{eq:mdot_tired}
\end{equation}
This is the ``photon-tired'' outflow rate\footnote{$\dot{M}_{\rm tir}$ does not usually include the $L_0-L_{\rm Edd}$ factor \citep[e.g.,][]{owocki_1997}. We opt for it because then $m=1$ is the maximum tiring parameter (rather than being some value less than $1$ that depends on $L_0$).}. The tiring parameter,
\begin{equation}
m\equiv\frac{\dot{M}_{\rm out}}{\dot{M}_{\rm tir}},
\end{equation}
tells us how photon-tired the wind is. If $m=1$, then $v_\infty=0$, and $\lim_{r\rightarrow\infty}L=L_{\rm Edd}$. 

Photon-tired winds were proposed to explain dramatic episodes of mass loss in some massive stars, most notably the 1837-1856 giant eruption of $\eta$ Carinae\footnote{Interestingly, light echoes from $\eta$ Carinae's 1837-1856 giant eruption feature spectra with similar temperatures as LRDs \citep[$\sim5000-7000\,{\rm K}$,][]{rest_prieto_2012,owocki_shaviv_2016}. This connection was recently explored by \cite{naidu_2026}.}  \citep{owocki_2004,owocki_2012}. The same physics may hold in a quasi-spherical wind launched from $r_{\rm sph}$ with outflow rate $\dot{M}_{\rm out}$. Then, the tiring parameter is
\begin{equation}
    m \approx 0.2\,\frac1{\Gamma_0-1}\left(\frac{\dot{M}_{\rm out}}{\dot{M}}\right)\left(\frac{h}{0.3}\right),
    \label{eq:tiring_parameter_redd}
\end{equation}
indicating that tiring is important if $\Gamma_0\gtrsim 1$ and $\dot{M}_{\rm out}\sim\mathcal{O}(\dot{M})$. Equation~\ref{eq:tiring_parameter_redd} has no dependence on the BH mass or accretion rate. This is not a coincidence; $r_{\rm sph}$ is where radiation only just supports the disk against gravity,  so $\Gamma_0\gtrsim 1$ is expected ($\Gamma_0$ may be larger in the polar funnel). Then, the winds would be marginally unbound, and naturally lead to the fountain flow described in Sec.~\ref{sec:fountain_flow}.

The governing equations for the wind are, in addition to Eq.~\ref{eq:energy_conservation}, 
\begin{align}
\label{eq:acceleration}    v\frac{dv}{dr} &=\frac{\kappa_{\rm es}}{4\pi r^2 c}L(r)-\frac{GM}{r^2}\\
\label{eq:diffusion}    \frac{dP_{\rm rad}}{dr} &= -\rho \frac{\kappa_{\rm es}}{4\pi r^2 c}L(r) \\
\label{eq:continuity} \dot{M}_{\rm out}&=4\pi r^2 \rho v
\end{align}
Here, Eq.~\ref{eq:acceleration} is the acceleration of the wind, Eq.~\ref{eq:diffusion} is the diffusion of radiation through the wind, and Eq.~\ref{eq:continuity} is mass conservation. 

We solve these equations, which reduce to a set of two coupled first-order ODEs, following Appendix A of \cite{owocki_2017} using a two-sided shooting method: we integrate outward from an inner boundary at $r_{\rm sph}$ and inward from an arbitrarily large outer boundary. Both endpoints are singular, so integrations begin a small distance inside each boundary using series expansions. We require the two solutions match at an intermediate radius (we use $2\,r_{\rm sph}$; the choice is arbitrary). As the tiring parameter approaches $m=1$, the terminal velocity approaches zero and the shooting method struggles. Then, we instead solve the equations globally with a collocation method, where we discretize the domain using a logarithmically-spaced grid. Here, we invert the problem, and guess for the terminal velocity  and treat the tiring parameter as an unknown. The system is closed by two conditions: (i) the velocity vanishes at $r_{\rm sph}$ (we found little difference when setting the velocity at $r_{\rm sph}$ to the local gas sound speed, which is negligible); and (ii) the radiation pressure vanishes at infinity. 
The code is publicly available at \url{https://github.com/nkaaz/owocki_winds}.

We show the density and velocity profiles of the photon-tired wind solutions in Fig.~\ref{fig:wind_profiles} for $\dot{M}_{\rm out}=2.5$, $5$, $10$ and $15\,M_\odot\,{\rm yr}^{-1}$ and $m=0.6$, $0.9$ and $0.94$. These values of $m$ correspond to $\Gamma_0\approx1.33$, $1.22$ and $1.21$, respectively (Eq.~\ref{eq:tiring_parameter_redd}). We do not plot the luminosity as a function of radius; in each model, the luminosity is reduced by $\sim17\%$. At $m=0.94$ and $0.9$, the scattering photosphere is at $r_{\rm sca}\sim1-5\times10^2\,r_{\rm sph}\sim 10^{17}-10^{18}\,{\rm cm}$ where the velocity is $\gtrsim100-500\,{\rm km\,s^{-1}}$. These values are consistent with LRDs. At $m=0.6$, the wind becomes too fast to explain LRDs. So, if LRD envelopes are photon-tired winds, then $m$ must be very close to unity. Near-unity tiring parameters may represent attractor solutions, wherein the light unbinds as much gas as it possibly can; this possibility should be explored in numerical simulations. From here on out, we focus only on the $m=0.94$ solutions.

\begin{table*}
  \centering
  \setlength{\tabcolsep}{0pt}
  \caption{Wind models used in \textsc{Sirocco} calculations. The SMBH mass is always $10^6\,M_\odot$. The input intensity at the inner boundary is $I_\nu=WB_\nu$, where $W=1/\tau_{\rm in}$ is the dilution factor. $T_{\rm bb}$ is set to normalize the total luminosity to $L_{\rm Edd}$ ($\Gamma_0L_{\rm Edd}$) in the marginally unbound (photon-tired) runs. We list the velocity at the scattering photosphere, $v_{\rm ph}$ for the marginally unbound winds, which continue to decelerate, and we list the asymptotic velocity, $v_\infty$, for the photon-tired winds. The inner boundary radius is $r_{\rm in}=r_{\rm sph}$ for the marginally unbound winds and $r_{\rm in}=\max[r_{\rm sph},\,r(\tau_{\rm in}\!=\!100)]$ for the photon-tired winds. Models are named \texttt{mu\_m}$X$ and \texttt{pt\_m}$X$, where $X$ is $\dot{M}_{\rm out}$ in $M_\odot\,{\rm yr^{-1}}$.}
  \label{tab:wind_models}
  \begin{tabular*}{\textwidth}{@{\extracolsep{\fill}}ccccccccc@{}}
  \hline\hline
  & \multicolumn{4}{c}{Marginally Unbound Winds} & \multicolumn{4}{c}{Photon-Tired Winds} \\
  \cline{2-5}\cline{6-9}
  $\dot{M}_{\rm out}\,[M_\odot\,{\rm yr^{-1}}]$ &
  $v_{\rm ph}\,[{\rm km\,s^{-1}}]$ & $r_{\rm in}\,[10^{15}\,{\rm cm}]$ & $T_{\rm bb}\,[10^4\,{\rm K}]$ & $W$ &
  $v_\infty\,[{\rm km\,s^{-1}}]$ & $r_{\rm in}\,[10^{15}\,{\rm cm}]$ & $T_{\rm bb}\,[10^4\,{\rm K}]$ & $W$ \\
  \hline
  2.5 & 311 & 0.80 & 4.9 & 0.053 & 306 & 2.07  & 4.9 & 0.010 \\
  5   & 156 & 1.60 & 3.8 & 0.037 & 217 & 5.38  & 3.0 & 0.010 \\
  10  & 78  & 3.20 & 2.9 & 0.026 & 153 & 14.2  & 1.9 & 0.010 \\
  15  & 52  & 4.80 & 2.5 & 0.022 & 125 & 25.1  & 1.4 & 0.010 \\
  \hline\hline
  \end{tabular*}
\end{table*}

%%%%%%%%%%%%%%%%%%%%%%%%%%%%%%%%%%%%
%% SECTION 3 : RADIATIVE TRANSFER %% 
%%%%%%%%%%%%%%%%%%%%%%%%%%%%%%%%%%%%

\section{Radiative Transfer Calculations}
\label{sec:radtransfer}

Do the slow, super-Eddington winds we have advocated for in Sec.~\ref{sec:outflow_problem}-\ref{sec:lrds_and_stars} manifest as LRDs? To answer this, we perform radiative transfer calculations of both the marginally unbound and photon-tired winds. The marginally unbound winds are ad hoc, but qualitatively capture the behavior of a fountain flow; the photon-tired winds are physically motivated, though model-dependent, and provide a plausible mechanism for driving such a flow. We will ultimately find that both tell a similar story.

\subsection{Numerical Details}
\label{sec:numerical}

\textit{Code info.} We calculate the SED and radiative structure of the marginally unbound and $m=0.94$ photon-tired winds by using the open-source Monte Carlo (MC) radiative transfer code \textsc{Sirocco}\footnote{We made minor modifications to \textsc{Sirocco} -- primarily the analysis scripts -- which are publicly available \href{https://github.com/nkaaz/LRDs-as-Super-Eddington-Fountain-Flows_REPRODUCIBLE}{here}.} \citep{sirocco_codepaper}. \textsc{Sirocco} samples photon packets from an input intensity distribution at the inner boundary of the domain that then propagate through a user-specified density and velocity field. The code accounts for special relativistic effects, includes inelastic electron scattering, and treats bound-bound absorption in the Sobolev \citep{sobolev_1957,CAK75} approximation (we discuss the Sobolev approximation and its limitations in Sec.~\ref{sec:caveats}). The scheme first undergoes ``ionization cycles'', which iteratively determine the ionization state and electron temperature of the gas in every cell, and then undergoes ``spectrum cycles'' which calculate the resulting spectrum in specified bands. 

%% FIGURE SPECTRA GALLERY
\begin{figure*}[bht]
    \centering
    \includegraphics[width=0.9\textwidth]{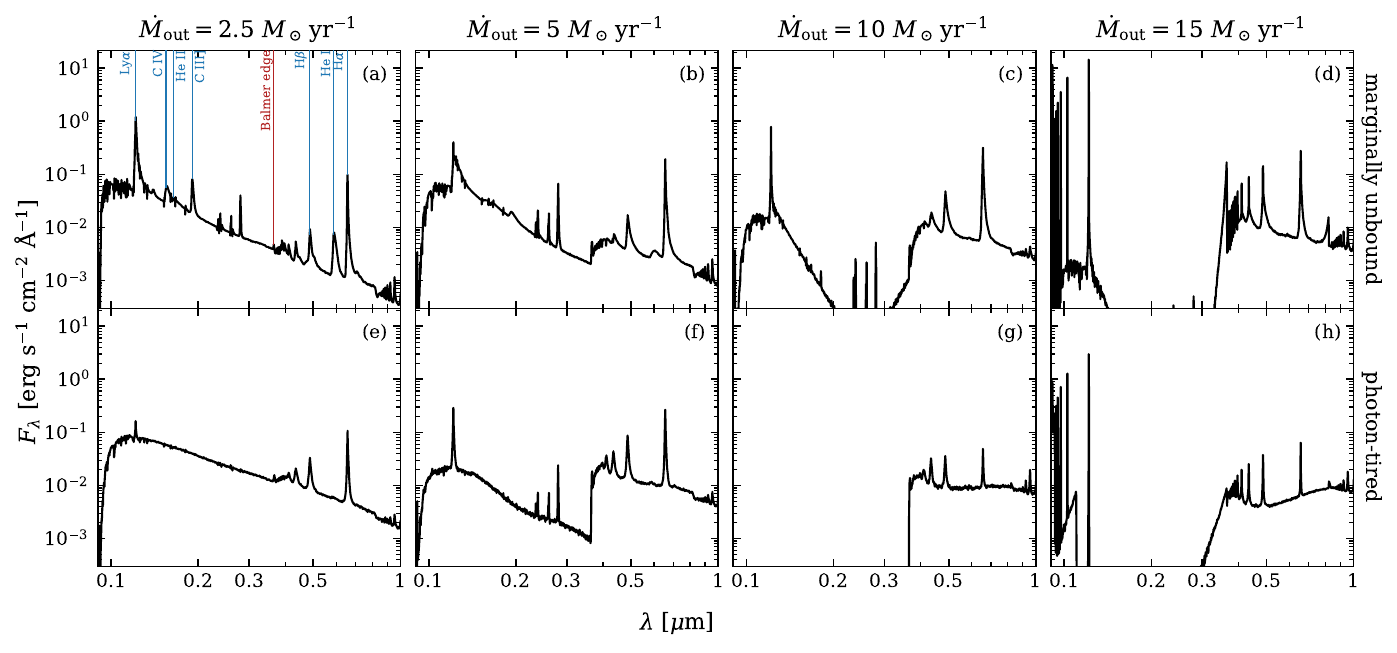}
    \caption{Synthetic spectra from each wind model. At lower mass loss rates, there is no Balmer break, and the wind appears as a ``little blue dot'' (panels a and e); at intermediate mass outflow rates, there is a partial Balmer break, and the wind appears as an LRD with an intrinsically V-shaped spectrum (panels b, c and f); at high mass loss rates, the Balmer break is complete, and kills all blue emission (panels d, g and h). Panels d and h use ``truncated'' spectra that excise outer, unconverged cells beyond the recombination front (see Sec.~\ref{sec:numerical} and App.~\ref{app:truncated_spectra_comparison}). Note that in panels d, g and h, the complete Balmer breaks result in a paucity of UV photons, such that Monte Carlo shot noise pollutes the spectra near $0.1\,\mu {\rm m}$. The depicted spectra are also smoothed over with a $1000\,{\rm km\,s^{-1}}$ boxcar.}
    \label{fig:spectra}
\end{figure*}
%% FIGURE HALPHA GALLERY
\begin{figure*}[bht]
    \centering
    \includegraphics[width=0.9\textwidth]{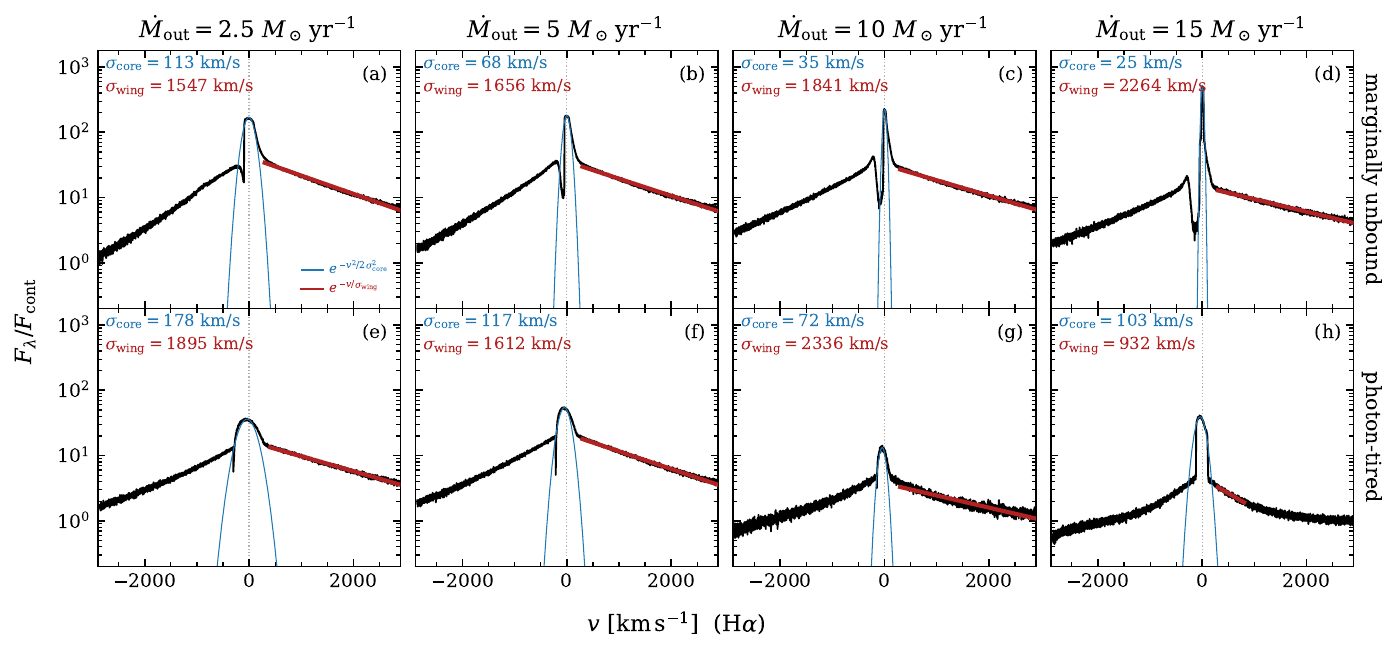}
    \caption{Synthetic H$\alpha$ profiles from each model (same as in Fig.~\ref{fig:spectra}). Electron scattering wings, narrow emission cores, and blueshifted absorption troughs appear in all spectra. We fit the red wings of the line profiles to exponentials and the narrow cores to Gaussians (labeled in panel a). As in Fig.~\ref{fig:spectra}, panels d and h use spectra from ``truncated'' runs that excise outer, unconverged cells beyond the recombination front (see Sec.~\ref{sec:numerical} and App.~\ref{app:truncated_spectra_comparison}).} 
    \label{fig:halpha}
\end{figure*}

\textit{Treatment of atoms.} \textsc{Sirocco} treats the atomic level populations and the reprocessing of absorbed radiation using either a ``simple-atom'' or ``macro-atom'' approximation. In the simple atoms, ionization and excitation are treated separately. The ionization state is set by a rate matrix in which photoionization and collisional ionization balance recombination; the photoionization rate is determined by fitting the local mean intensity to a piecewise power-law or exponential (\textsc{Sirocco}'s \texttt{matrix\_pow} mode). Then, given the determined neutral fraction, each line is treated as an isolated two-level system. The lower-level population is determined by a Boltzmann distribution at the local radiation temperature and dilution factor, and the upper-level population follows from local statistical equilibrium (collisional rates plus radiative pumping by the ambient field). In simple atoms, the level populations are not solved self-consistently -- a given level is assigned different populations depending on whether it enters a transition as the upper level (from statistical equilibrium) or the lower level (from the Boltzmann distribution). The simple-atom treatment permits no radiative cascades through intermediate levels: a photon absorbed in a line is either re-emitted in the same transition or collisionally thermalized, and recombination contributes only continuum emission (precluding both fluorescence and recombination cascades).

The macro-atom approximation instead solves for the level populations self-consistently under statistical equilibrium, using estimates of the radiation field together with all radiative and collisional rates, and chooses the fate of an absorbed packet among all interaction pathways. We adopt the more expensive macro-atom treatment for hydrogen and helium and the two-level simple-atom treatment for the metals (C, N, O, Ne, Na, Mg, Al, Si, S, Ar, Ca, and Fe). Hydrogen is modeled as a 20-level macro-atom and helium as a 53-level (He I) plus 10-level (He II) macro-atom. We use the \texttt{h20\_hetop\_standard80} line list distributed with \textsc{Sirocco}, which includes roughly 5,400 metal bound-bound transitions. The hydrogen and He II macro-atom levels are resolved by principal quantum number n (up to n = 20 for hydrogen), with the orbital angular momentum (l) and fine-structure sublevels collapsed into a single level. This assumes the sublevels are populated in proportion to their statistical weights, which holds when l-changing collisions are fast enough to maintain statistical equilibrium -- i.e., at sufficiently high density\footnote{Because the l-sublevels of a given n of a hydrogenic atom are nearly degenerate, only a little energy is needed to move between them; l-changing collisions can therefore remain rapid even where collisional excitation and ionization are entirely negligible.}. He I is instead resolved by $n$, $l$ and spin, with $n\leq10$ and $l\leq2$.  In the low density outer wind this assumption may break down, and an l-resolved treatment should be considered in future work \citep{pengelly_1964,draine_ISM}.

\textit{Problem Setup.} We use a spherically symmetric domain with, usually, $100$
logarithmically spaced radial cells. In \texttt{pt\_m15}, we used $300$ cells instead to improve    convergence. Photon packets are injected at the inner
boundary, $r_{\rm in}$, and sampled from the distribution $I_\nu = W B_\nu(T_{\rm bb})$.
We usually set $r_{\rm in}=r_{\rm sph}$; in the photon-tired wind runs it must be
placed further out, because it is difficult for the Monte Carlo radiation scheme
to propagate photons through the inner density cusp (see Fig.~\ref{fig:wind_profiles}b). In these cases we instead choose $r_{\rm in}$ by capping
the inner optical depth at $\tau_{\rm in}=100$, where $\tau_{\rm in}=\int_{r_{\rm in}}^{r_{\rm out}}\kappa_{\rm es}\,\rho\,\mathrm{d}r$
is the electron-scattering optical depth between the inner and outer boundaries for a fully ionized medium. We set the dilution factor to $W=1/\tau_{\rm in}$ and the blackbody temperature
$T_{\rm bb}$ such that $L=L_{\rm Edd}$ for the marginally unbound runs and $L=\Gamma_0L_{\rm Edd}$ for the photon-tired runs\footnote{Note, that in the photon-tired wind models described in Sec.~\ref{sec:photon_tired_winds}, some of the luminosity is depleted at radii $r_{\rm sph}<r<r_{\rm in}$. This results in a $\lesssim17\%$ discrepancy with the injected luminosity and we do not expect that it affects our results.}. Here, we use a dilution factor to capture the trapping of radiation at the base of the optically thick wind, and $W=1/\tau_{\rm in}$ estimates the standard gray-diffusion result. We use a reflecting boundary condition at $r_{\rm in}$. We set $r_{\rm out}$ to at least ten times the electron scattering photosphere ($\tau_{\rm es}=1$, assuming fully ionized gas). We adopt $10\%$ solar metallicity for all calculations and include adiabatic cooling of the wind.

% \nko{replace this when i change to matrix\_pow; perhaps say less and emphasize that this only affects two-level atoms}. We solve for the ionization state of each cell with \textsc{Sirocco}'s matrix
% scheme (\texttt{matrix\_bb} in the code), which determines the ionization
% fractions by balancing the full set of photoionization, collisional, and
% recombination rates through direct inversion of the rate matrix, rather than the
% simpler on-the-spot or modified nebular approximations. The radiation field
% entering the photoionization rate estimators is modelled as a dilute blackbody,
% parametrized by a radiation temperature and dilution factor fit to
% the Monte Carlo mean intensity in each cell. 

We use
$2\times10^6$ photon packets per cycle, with $14$ ionization
cycles and, usually, $20$--$40$ spectrum cycles. The emergent spectra are computed from $200$ to $20{,}000\,\text{\AA}$ on a uniform grid
of $2\times10^{5}$ log-wavelength bins,
corresponding to a velocity resolution of $6.9\ \mathrm{km\,s^{-1}}$. We additionally recompute five spectrum cycles over a narrow window, $6500$--$6700\,\text{\AA}$ with $10^{4}$ bins, giving a velocity resolution of
$\approx\!0.9\ \mathrm{km\,s^{-1}}$ at H$\alpha$.

In runs \texttt{mu\_m15} and \texttt{pt\_m15}, strong recombination fronts form, beyond which the calculations are poorly converged. The density in this region artificially reprocesses their spectra. For this reason, we performed ``truncated'' calculations of these two runs, where we excised the gas  beyond $8.5\times10^{18}\,{\rm cm}$ and $5.2\times10^{17}\,{\rm cm}$ for \texttt{mu\_m15} and \texttt{pt\_m15}, respectively (before, $r_{\rm out}=10^{20}\,{\rm cm}$ and $2\times10^{19}\,{\rm cm}$). The truncated calculations are used only for the spectra in Figs.~\ref{fig:spectra}-\ref{fig:halpha}; we compare the spectra of the truncated versus fiducial calculations in App.~\ref{app:truncated_spectra_comparison}. Overall, we find that the truncated spectra miss certain absorption features in the spectra, but do not qualitatively change our main results. 

The density and velocity profiles for the two families are given in
Sec.~\ref{sec:fountain_flow} and Sec.~\ref{sec:photon_tired_winds}.
For the photon-tired winds we specialize to $m=0.94$, which produce the slowest winds and are most representative of LRDs (see
Fig.~\ref{fig:wind_profiles}a). Each run is listed in Table~\ref{tab:wind_models}.

\subsection{Spectra}
\label{sec:spectra}

In Fig.~\ref{fig:spectra}, we show the continuum spectra of each marginally unbound wind (top row) and each photon-tired wind (bottom row). $\dot{M}_{\rm out}$ increases from left to right. In the first column, $\dot{M}_{\rm out}= 2.5\,M_\odot\,{\rm yr}^{-1}$, and the wind is not dense enough for a Balmer break to form.  We interpret these as ``little blue dots'' (LBDs), an emerging class of AGN that lack a Balmer break but share other LRD hallmarks \citep[X-ray \& radio weakness, exponential emission line wings, lack of variability; see][]{brazzini_2026}. In this interpretation, LBDs are powered the same way as LRDs \citep[as argued by][]{sneppen_2026_lbd, madau_maiolino_2026}, except that the winds are less dense along our line of sight. This may occur either at viewing angles closer to face-on (where the wind is faster, and thus less dense at the same outflow rate, Sec.~\ref{sec:outflow_problem}) or at the same off-axis viewing angle as an LRD but at lower overall outflow rates.

In the second column of Fig.~\ref{fig:spectra}, $\dot{M}_{\rm out}=5\,M_\odot\,{\rm yr}^{-1}$. The marginally unbound wind features about a factor of two drop at the Balmer edge. Partial Balmer breaks result in a `V-shaped' SED, typical of LRDs \citep{setton_2025_vshape}, where blue emission from the AGN remains important. In the third column of Fig.~\ref{fig:spectra}, $\dot{M}_{\rm out}=10\,M_\odot\,{\rm yr}^{-1}$, and the wind is opaque to Balmer-edge photons. This results in a strong break at $0.365\,\mu$m, but allows bluer photons to escape (the bound-free opacity scales as $\nu^{-3}$). In the fourth column of Fig.~\ref{fig:spectra}, $\dot{M}_{\rm out}=15\,M_\odot\,{\rm yr}^{-1}$, and the wind is completely opaque to the Balmer continuum ($3.4\,{\rm eV}<h\nu<13.6\,{\rm eV}$), preventing blue light from escaping (the far-UV spikes are polluted by Monte Carlo shot noise). This is observed in a subset of LRDs \citep[e.g.,][]{naidu_2025,degraaff_2025}. For a direct comparison with observed systems, we refer the reader to \cite{sneppen_2026a}, who performed similar calculations.

We also find Paschen jumps at $\dot{M}_{\rm out}\lesssim10-15\,M_\odot\,{\rm yr}^{-1}$. This is consistent with the \textsc{Sirocco} calculations by \cite{sneppen_2026_paschen}, who also found evidence for Paschen jumps in observed LRDs \citep[including the ``Rosetta Stone'' LRD GN-28074,][]{juodzbalis_2024}. Interestingly, in the $\dot{M}_{\rm out}=15\,M_\odot\,{\rm yr}^{-1}$ photon-tired wind, we find a V-shape at $\lambda\sim0.5\,\mu {\rm m}$. This is possibly due to Paschen absorption or absorption in the Balmer-line forest. This hints at an upper limit on the outflow rate in LRDs, given that this feature has not (yet) been observed (we discuss Paschen jumps and breaks further in Sec.~\ref{sec:photoionization}). However, this result may depend sensitively on the Paschen optical depth, which is set by Balmer resonant scattering. Given the uncertainties in line transfer under the Sobolev approximation (see Sec.~\ref{sec:caveats}), this result should be taken with a grain of salt. 

%% FIGURE ELECTRON TEMP AND IONIZATION FRACTION 2 PANEL SINGLE COLUMN
\begin{figure}%[bht]
    \centering
    \includegraphics[width=0.9\textwidth]{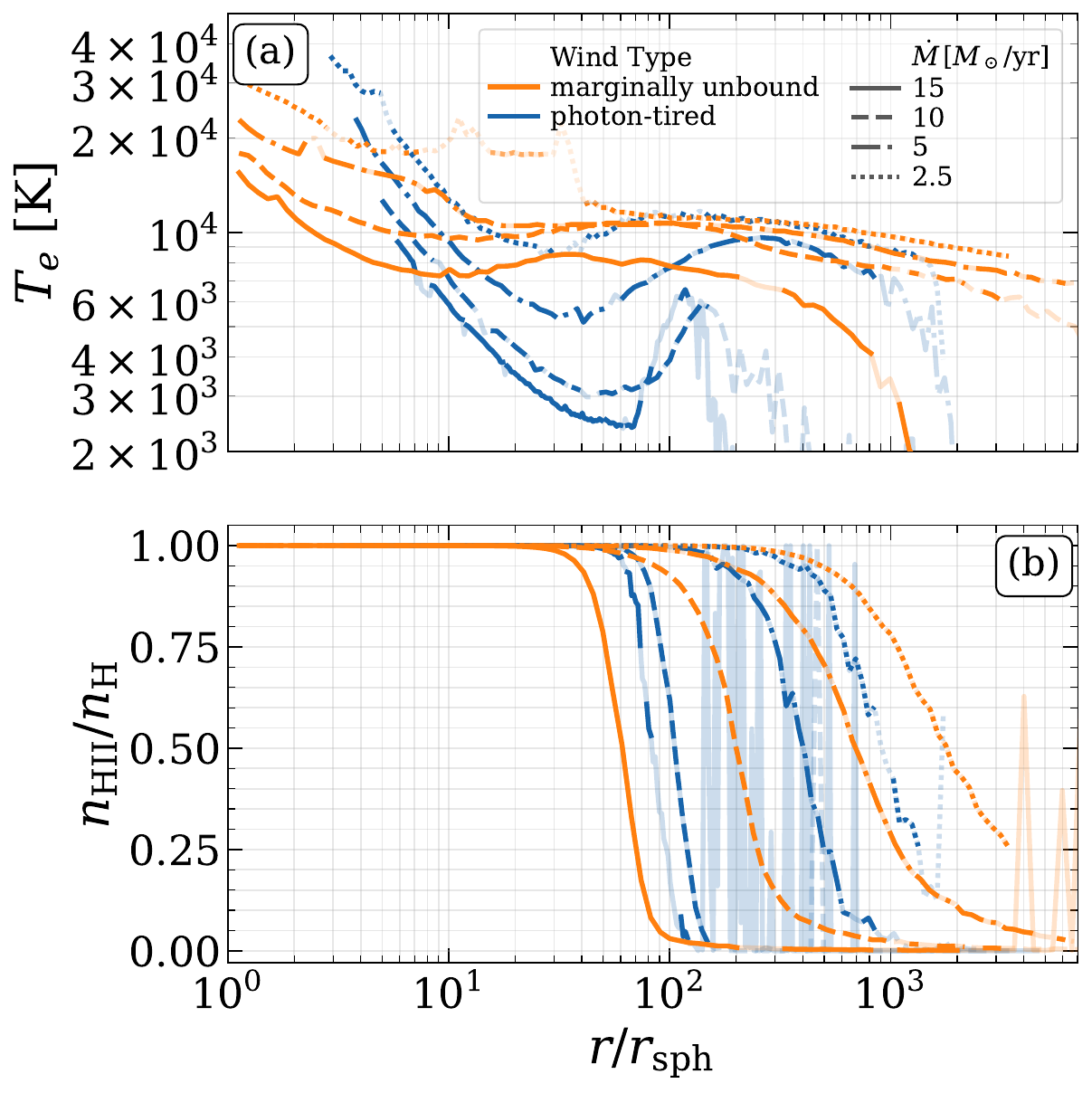}
    \caption{Electron temperature and ionization state of each wind. Invariably, the electron temperature first lowers from $\sim1-4\times10^4\,{\rm K}$ before plateauing at $T_e\lesssim10^4\,{\rm K}$. The wind recombines around $r\gtrsim10^2\,r_{\rm sph}\gtrsim 10^{17}\,{\rm cm}$, either because the wind becomes optically thin and the radiation field can no longer photoionize it, or because the photoionizing continuum is depleted and the gas recombines. The faded line segments have poor convergence.} 
    \label{fig:temp_ion}
\end{figure}

%% FIGURE DETAILED BALANCE FULL PAGE
\begin{figure*}
    \centering
    \includegraphics[width=0.9\textwidth]{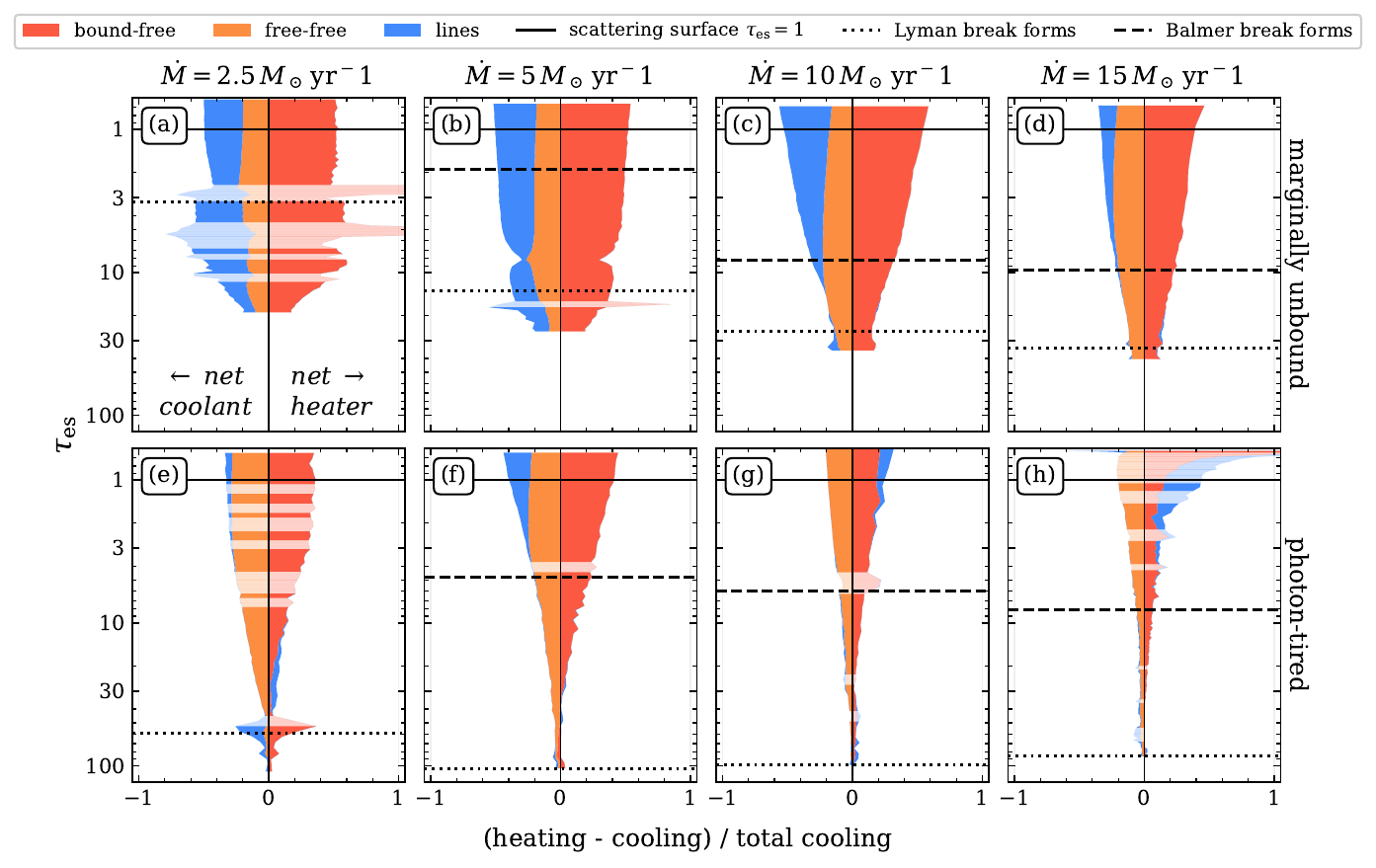}
    \caption{Balance of heating and cooling for various processes as a function of the scattering optical depth, which is a proxy for radius. We only plot $\tau_{\rm es}>0.6$ and account for ionization fraction. We show the aggregate of all bound-free, free-free and line opacities. Each colored region's width is the net (heating - cooling) of that process, normalized to the total cooling rate (summed over all processes). Wherever the heating and cooling for a given process sum to zero, it is in detailed balance. The larger the deviation from zero, the farther from LTE the gas is. We mark the scattering photosphere (solid line), the location of the Lyman break (dotted line), and (if present) the location of the (full) Balmer break (dashed line). Throughout, the gas is near LTE where $\tau_{\rm es}\gtrsim20-30$, and departs beyond, with bound-free absorption usually being the main process heating the gas. The gas is closer to achieving LTE when $\dot{M}_{\rm out}$ is higher, but is always out of LTE close to the photosphere. Note that these are aggregate sums over all bound-free processes; for instance, Lyman and Balmer photoionization heat the gas while all remaining processes usually cool the gas, which this plot does not capture. The washed-out shaded regions have poor convergence.}
    \label{fig:detailed_balance}
\end{figure*}

In Fig.~\ref{fig:halpha}, we show the H$\alpha$ emission line profiles. The line profiles show three characteristic features: a narrow emission core of width $\sim10^2\,{\rm km\,s^{-1}}$, an absorption trough blueshifted by a similar velocity, and broad exponential wings. The narrow core and the blueshifted absorption trough together form a ``P~Cygni'' line profile. The narrow core is produced near the scattering photosphere; it is symmetric because we detect photons emitted from both near (blueshifted) and far (redshifted) ends of the wind. The width of the narrow core is set by the velocity at the photosphere. The blueshifted absorption feature occurs because some of the line photons are scattered out of our line of sight. The wings are $\sim1000-2000\,{\rm km\,s^{-1}}$ wide and are well described by exponential profiles. The exponential shape is caused by repeated scatters that transfer energy between electrons and photons \citep{laor_2006}. Here, the scattering wings are asymmetric because the envelope is idealized to be one-dimensional and entirely outflowing; in nature, we instead expect a fountain flow that has a mixture of inflowing and outflowing streamlines (see Sec.~\ref{sec:fountain_flow} and Fig.~\ref{fig:cartoon}) that may produce more symmetric profiles \citep{sneppen_2026a}. 

%% FIGURE HA HB
\begin{figure}%[bht]
    \centering
    \includegraphics[width=\textwidth]{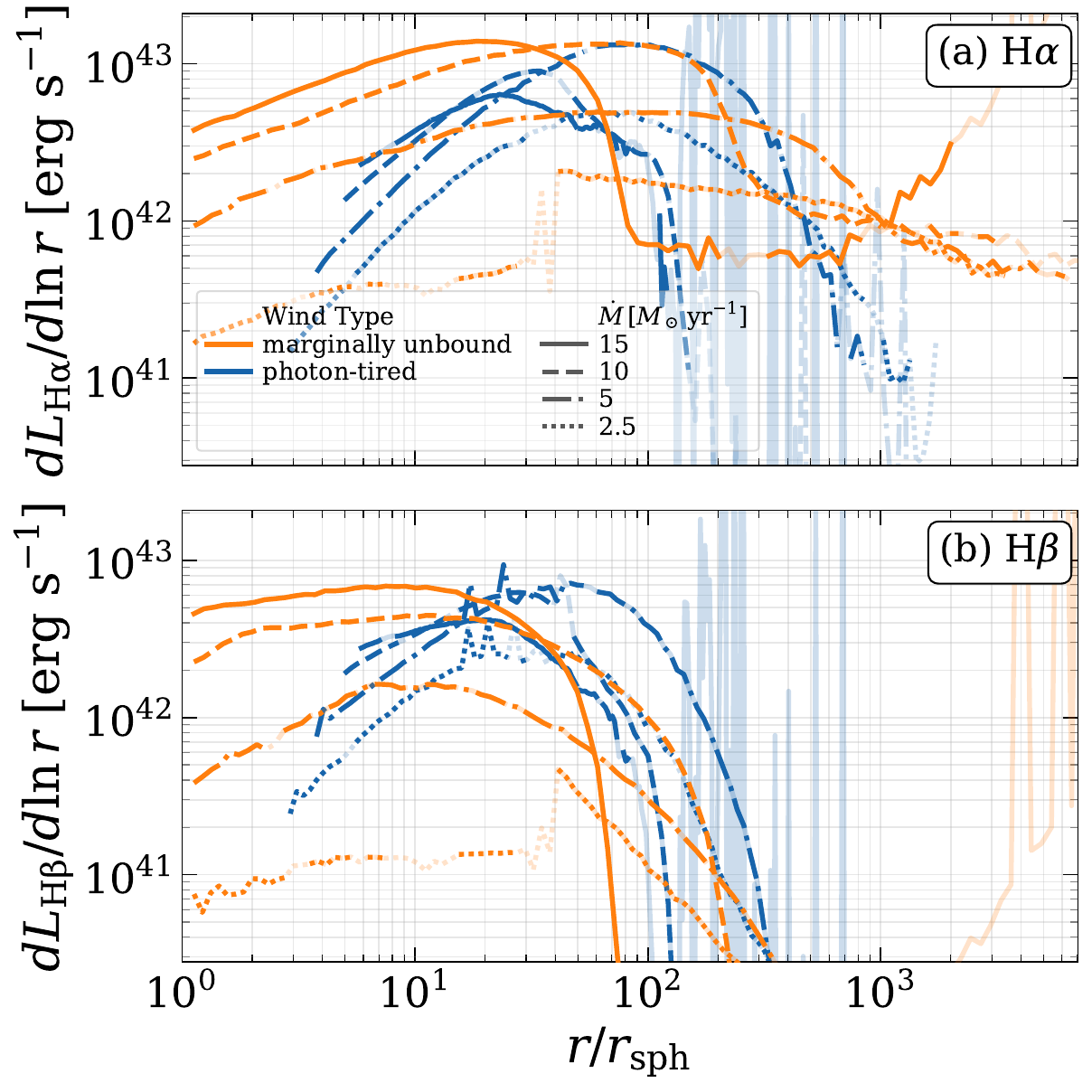}
    \caption{Where H$\alpha$ and H$\beta$ Balmer lines are formed. We plot the net line emission per log radius (Eq.~\ref{eq:line_emission_function}) for H$\alpha$ (top) and H$\beta$ (bottom), with the Sobolev escape probability $\beta_{\rm esc}$ suppressing trapped emission deep in the wind. The faded segments have poor convergence.}
    \label{fig:net_line_emissivity}
\end{figure}

 %% FIGURE OPTICAL DEPTH FULL PAGE
\begin{figure*}
    \centering
    \includegraphics[width=0.9\textwidth]{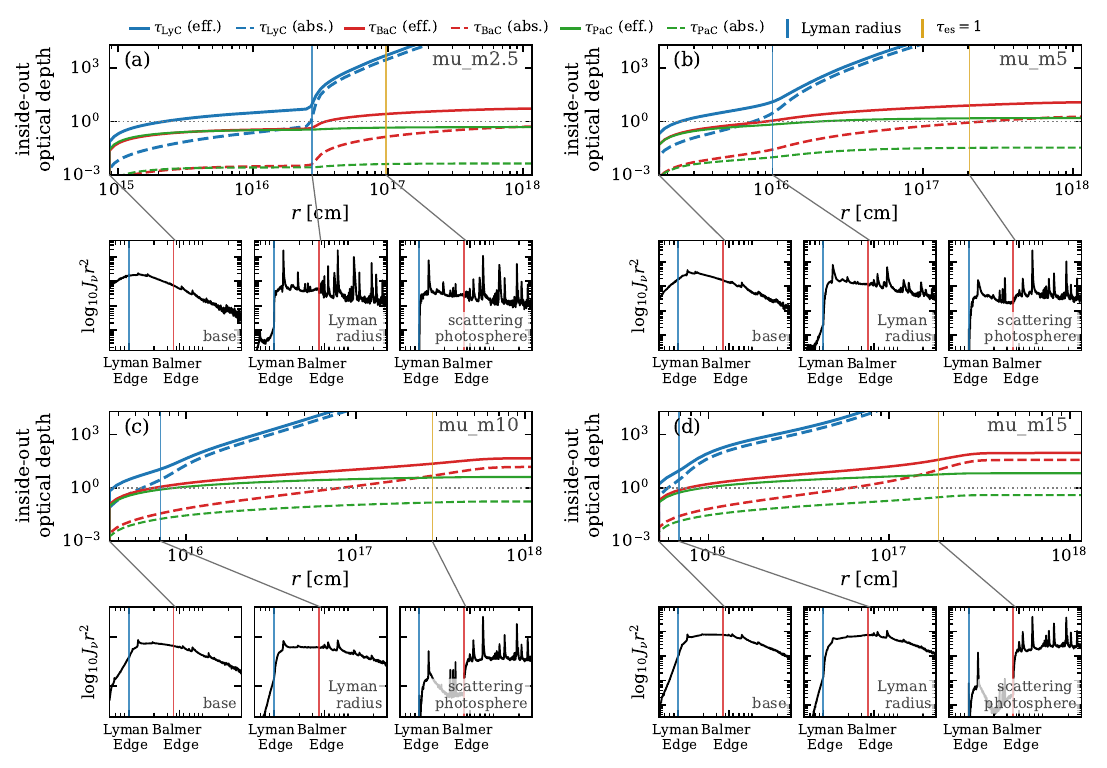}
    \caption{Optical depth as a function of radius in the marginally unbound wind \textsc{Sirocco} calculations, including inset panels of the mean intensity $J_\nu r^2$ at key radii. In panels (a)-(d), we show both the absorption and effective optical depths for Lyman, Balmer and Paschen bound-free opacities, evaluated at each bound-free edge. In all runs, a Lyman break forms within the wind, beyond which the Lyman continuum is fully reprocessed. We label this with a blue vertical line at $\tau_{\rm LyC}=3$ and refer to it later as the ``Lyman transition radius''. At higher $\dot{M}_{\rm out}$, Balmer absorption becomes more and more important, forming strong Balmer breaks. Paschen absorption is also marginally important.
    \label{fig:optical_depth}}
\end{figure*}

\subsection{Radiative Structure}

The electron temperature, $T_e$, shown in Fig.~\ref{fig:temp_ion}a, is $\gtrsim2-4\times10^4\,{
\rm K}$ near $r_{\rm sph}$. The gas cools with radius until $T_e$ plateaus to $\lesssim10^4\,{\rm K}$. Up until about $\gtrsim10^2\,r_{\rm sph}$, the wind remains fully ionized; we show the ionization fraction in Fig.~\ref{fig:temp_ion}b. Beyond, the gas recombines. The gas is primarily heated by photoionization and is out of LTE except near the base of the wind. 

At our densities  ($\lesssim 10^{11}\,{\rm cm}^{-3}$, usually), we expect that collisions are unimportant for determining the level populations of hydrogen. However, we caution that we have not definitely established this (though, Govreen-Segal et al. (in prep) has ran similar calculations without collisions, and found similar results). If collisions are unimportant, LTE can only be established radiatively, which requires the radiative processes to be in detailed balance. Detailed balance in turn requires the heating and cooling rates of each radiative process to balance independently. \textsc{Sirocco} outputs the heating and cooling rate of the bound-free, free-free and line processes. In Fig.~\ref{fig:detailed_balance}, 
we plot the heating minus cooling rate of each radiative process as a function of the electron scattering optical depth, 
\begin{equation}
     \tau_{\rm es}(r) = \int_r^{r_{\rm out}} n_e \sigma_T dr,
\end{equation}
and normalize by the total cooling rate summed over all processes. In detailed balance, every curve would lie at zero; nonzero values indicate that a process is out of balance. When $\tau_{\rm es}\gtrsim30$, the radiative processes are nearly balanced. Closer to the photosphere, bound-free absorption is the dominant heating mechanism, while free-free and (usually) line emission cools the gas. At higher $\dot{M}_{\rm out}$, the gas is closer to LTE; there may be a threshold $\dot{M}_{\rm out}$ above which LTE becomes a good approximation. Note that because we measure each process type in aggregate, Fig.~\ref{fig:detailed_balance} may over-estimate how close the radiative processes are to detailed balance; heating and cooling sub-channels of opposite sign partially cancel within each curve (for instance, Lyman and Balmer bound-free may cause net heating whereas Paschen and Brackett bound-free may cause net cooling; see also Sec.~\ref{sec:photoionization}).   

In Figure~\ref{fig:net_line_emissivity}, we explore where the H$\alpha$ and H$\beta$ emission lines are formed by calculating the net line emission rate per log radius,
\begin{equation}
\begin{aligned}
    \frac{dL_{ul}}{d{\rm ln}r}&= 4\pi r^3 n_u A_{ul}h\nu_{ul}\beta_{\rm esc}(\tau_S)\\&=4\pi r^3 n_u A_{ul}h\nu_{ul} (1-e^{-\tau_S})/\tau_S,
    \label{eq:line_emission_function}
\end{aligned}
\end{equation}
where $u$ and $l$ are the upper and lower levels of the line. Contributions from line photons at high optical depth are suppressed by the local escape probability $\beta_{\rm esc}$. \textsc{Sirocco} uses the Sobolev approximation (see Sec.~\ref{sec:caveats} for a discussion), where the optical depth is purely local \citep{hubeny_mihalas_2014},
\begin{equation}
    \tau_S=(\pi e^2/m_e c) f_{lu}c\nu_{ul}^{-1}[n_l-(g_l/g_u)n_u]/(|\mathrm{d}v/\mathrm{d}s|),
\label{eq:tau_s}
\end{equation}
For Balmer lines, $l=2$, and for H$\alpha$ (H$\beta$), $u=3$ ($u=4$). Here, the velocity gradient along the path length is $|\mathrm{d}v/\mathrm{d}s|\sim v/r$. Both lines are continuously produced from $10\lesssim r/r_{\rm sph}\lesssim10^2$. The photosphere is at $\gtrsim 10^2\,r_{\rm sph}$, indicating that most of the escaping line emission will be reprocessed by electron scattering (as evidenced by the broad exponential wings shown in Fig.~\ref{fig:halpha}). The oscillator strength, $f_{lu}$, for H$\alpha$ is $\approx5.4$ times larger than for H$\beta$, making H$\alpha$ more trapped.  

Next, we explore the key bound-free opacities. Lyman, Balmer and Paschen optical depths are (at their bound-free edges), 
\begin{align}    \label{eq:tau_lyc}\tau_{\rm LyC}(r) &= \int_{r_{\rm in}}^r n_1(r) \sigma_1 dr \\ 
\label{eq:tau_bac}\tau_{\rm BaC}(r) &= \int_{r_{\rm in}}^r n_2(r) \sigma_2 dr \\
\label{eq:tau_pac}\tau_{\rm PaC}(r) &= \int_{r_{\rm in}}^r n_3(r) \sigma_3 dr 
\end{align}
where $n_1$, $n_2$ and $n_3$ are the neutral hydrogen number densities in the $n=1$, $2$ and $3$ states, respectively. The bound-free cross-sections are $\sigma_1=6.3\times10^{-18}\,{\rm cm}^2$, $\sigma_2=1.39\times10^{-17}\,{\rm cm}^2$ and $\sigma_3=2.15\times10^{-17}\,{\rm cm}^2$. Note that we integrate these optical depths from the inside out. We do this because we want to know where a given population of photons (e.g., the Lyman continuum, $h\nu>13.6\,{\rm eV}$) emitted from $r_{\rm in}$ is absorbed. This calculation only makes sense if the base is not thermalized (as in the photon-tired winds, see Fig.~\ref{fig:detailed_balance}), where we would instead need to begin the integral at the thermalization radius. In practice, however, we only calculate the inside-out optical depth for the marginally unbound runs, where the base is not thermalized. 

Electron scattering also plays an important role in photoionization, because it traps the radiation and allows it to couple to the gas more readily. The effective optical depth is,
\begin{equation}
\tau_{\rm eff}=\sqrt{3\tau_{\rm abs}(\tau_{\rm es}+\tau_{\rm abs})},
\end{equation}
where $\tau_{\rm abs}$ is a stand-in for any absorption optical depth (as in Eqs.~\ref{eq:tau_lyc}-\ref{eq:tau_pac}, we also integrate from the inside out when calculating $\tau_{\rm eff}$). 

We plot the Lyman, Balmer and Paschen absorption and effective optical depths for each of the marginally unbound wind models in Fig.~\ref{fig:optical_depth} (the results do not meaningfully differ in the photon-tired winds). We also show inset panels of the mean intensity $J_\nu$ at key radii. In each wind, a strong Lyman break forms. We refer to this as the ``Lyman radius'' and define it as where $\tau_{\rm LyC}=3$. When $\dot{M}_{\rm out}=2.5\,M_\odot\,{\rm yr}^{-1}$ (panel a), the density of the wind is low enough that before the Lyman radius, the medium is optically thin to the Balmer photons. However, at the Lyman radius, the opacity gradient is steep. This is because the hydrogen population starts to recombine, increasing the neutral population, until Balmer photoionization becomes important. Still, in this run, the densities are low enough that no Balmer break forms.

Similarly, in the runs where  $\dot{M}_{\rm out}=5\,M_\odot\,{\rm yr}^{-1}$ (Fig.~\ref{fig:optical_depth}b) and $\dot{M}_{\rm out}=10\,M_\odot\,{\rm yr}^{-1}$ (Fig.~\ref{fig:optical_depth}c), the effective optical depth to the Balmer continuum crosses unity at the Lyman radius. When $\dot{M}_{\rm out}=15\,M_\odot\,{\rm yr}^{-1}$ (Fig.~\ref{fig:optical_depth}d), the Lyman radius is near $r_{\rm in}$. This is because the Lyman continuum is deeper in Wien's tail of the blackbody distribution (at higher $\dot{M}_{\rm out}$, $T_{\rm bb}$ is lower; see Table~\ref{tab:wind_models}), so there are fewer photons to deplete. The higher $\dot{M}_{\rm out}$ winds are more opaque to the Balmer photons, resulting in a Balmer break that scales with $\dot{M}_{\rm out}$, as can be seen by the inset panels showing $J_\nu r^2$ at the scattering photosphere. The location of the electron scattering photosphere is always $\gtrsim 10^{17}\, {\rm cm}$; it grows larger in radius until it peaks at $\sim3\times10^{17}\, {\rm cm}$ when $\dot{M}_{\rm out}=10\,M_\odot\,{\rm yr}^{-1}$, and reduces to $\lesssim2\times10^{17}\,{\rm cm}$ when $\dot{M}_{\rm out}=15\,M_\odot\,{\rm yr}^{-1}$. The photosphere is reached sooner because at high $\dot{M}_{\rm out}$, the Balmer continuum depletes (we will show this shortly), causing the gas to recombine.

%% FIGURE GAMMA1
\begin{figure}%[bht]
    \centering
    \includegraphics[width=\textwidth]{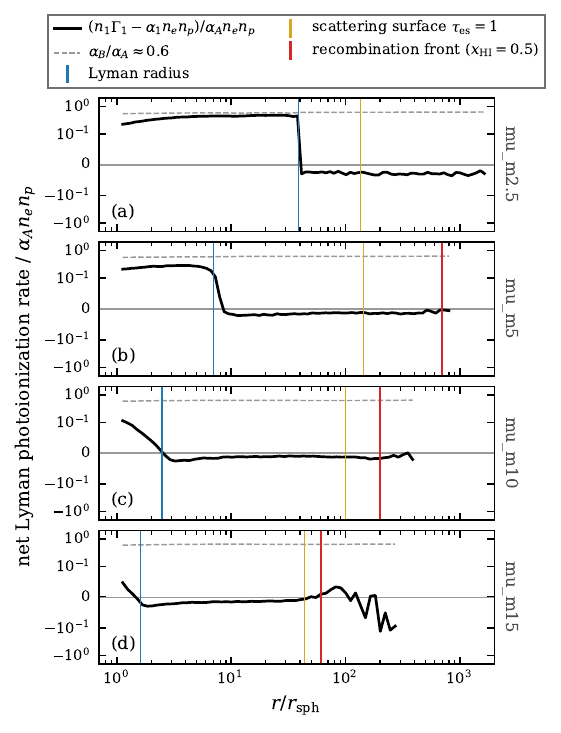}
    \caption{Net Lyman photoionization rate ($n_1\Gamma_1-\alpha_1 n_e n_p$) normalized to the total recombination rate ($\alpha_A n_e n_p$). The dashed line shows the limiting case, wherein the Lyman continuum is responsible for ionizing all recombined atoms. The Lyman continuum flux is depleted (replenished) where this quantity is positive (negative). It is always positive before the Lyman transition radius (blue vertical line), where the entire Lyman continuum flux is depleted. The net Lyman photoionization rate is near zero past the Lyman transition radius, indicating that the only remaining Lyman continuum photons are those produced due to recombinations directly to the ground state, which are exactly compensated for by photoionization. We also show the scattering surface with a yellow vertical line and the recombination front with a red vertical line.}
    \label{fig:gamma_lyc_resid}
\end{figure}

%% FIGURE GAMMA2
\begin{figure}%[bht]
    \centering
    \includegraphics[width=\textwidth]{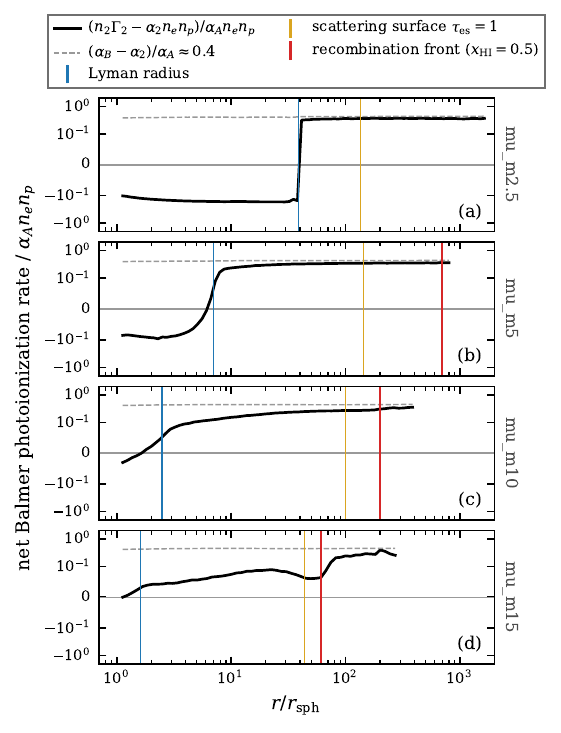}
    \caption{Same as Fig.~\ref{fig:gamma_lyc_resid}, except we show the net Balmer photoionization rate ($n_2\Gamma_2-\alpha_2 n_e n_p$) normalized to the total recombination rate ($\alpha_A n_e n_p$). In panels a-b, this quantity is negative before the Lyman transition radius. This indicates the net production of the Balmer continuum photons (this is due to the degradation of Lyman continuum photons, see Sec.~\ref{sec:photoionization}). Beyond the Lyman transition radius, the net Balmer photoionization rate is near the limiting value (dashed line), indicating that the Balmer continuum flux photoionizes all atoms that recombined to levels above the ground state. In panels c-d, the net Balmer photoionization rate is instead positive but below the limiting value, indicating that some of the atoms in higher excited levels are photoionized by other photoionization bands (e.g., Paschen) before they are able to cascade to lower levels.}
    \label{fig:gamma_bac_resid}
\end{figure}

%% FIGURE GAMMA3
\begin{figure}%[bht]
    \centering
    \includegraphics[width=\textwidth]{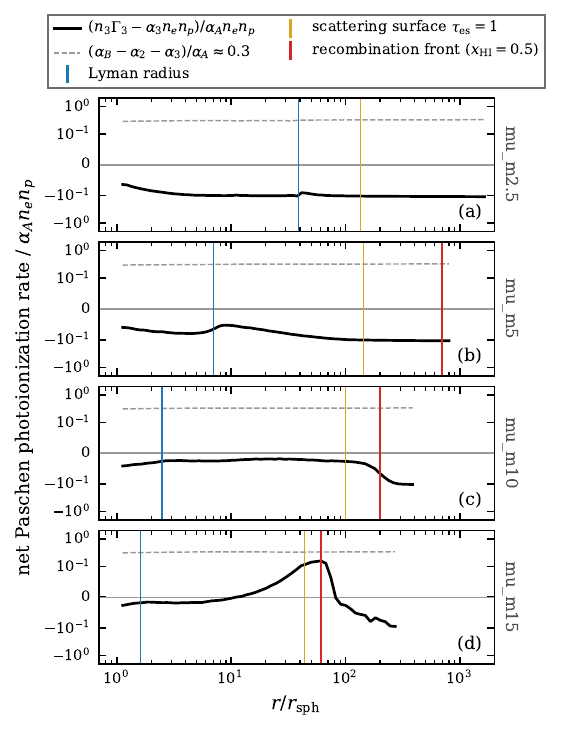}
    \caption{Same as Figs.~\ref{fig:gamma_lyc_resid}-\ref{fig:gamma_bac_resid}, except we show the net Paschen photoionization rate ($n_3\Gamma_3-\alpha_3 n_e n_p$) normalized to the total recombination rate ($\alpha_A n_e n_p$). This quantity is usually negative, indicating the net production of Paschen photons (this is due to photon degradation, see Sec.~\ref{sec:photoionization}). It is, however, positive and near its limiting value just beyond the scattering surface in panel d. Here, the Balmer continuum flux has been depleted, but Paschen photons keep the wind ionized slightly farther.}
    \label{fig:gamma_pac_resid}
\end{figure}

\subsection{Photoionization of the Wind}
\label{sec:photoionization}
Here, we investigate how the wind is photoionized in the \textsc{Sirocco} models. It is useful to write the equilibrium equation for the electron number density ($n_e$), 
\begin{equation}
    \sum_{i=1}^\infty [n_i\Gamma_i-n_en_p\alpha_i]=0,
    \label{eq:electron_density_equilibrium}
\end{equation}
where we have neglected collisions, the advection of electrons, and other atomic species. Here, $n_i$ is the number density of hydrogen in energy level $i$, $n_p$ is the ion number density, and $\alpha_i$ is the rate of recombinations to level $i$. The photoionization rate at level $i$ is,
\begin{equation}
    \Gamma_i = \int_{\nu_i}^\infty \frac{4\pi J_\nu}{h\nu}\sigma_i(\nu_i/\nu)^3 d\nu,
    \label{eq:photoionization_rate}
\end{equation}
where $\sigma_i$ is the absorption cross-section at each bound-free edge. Now, it is useful to do some ``photon accounting''. If hydrogen recombines directly to the ground state, it produces a photon above $13.6\,{\rm eV}$. If the same ground state atom is then photoionized, a photon above $13.6\,{\rm eV}$ is destroyed. So, if ever $n_1\Gamma_1=n_en_p\alpha_1$, there is no \textit{net} creation or destruction of Lyman continuum photons. 

Consider first the limiting case wherein the medium is solely photoionized by the Lyman continuum (as in, e.g., a Stromgren sphere). Then\footnote{Here, it is implicit that any recombination line beginning at level $i>1$ escapes; e.g., a recombination to $n=2$ first releases a Balmer continuum photon and then cascades to $n=1$, releasing an Lyman~$\alpha$ photon. }, 
\begin{equation}
    n_1\Gamma_1 = n_en_p \sum_{i=1}^\infty \alpha_i = n_e n_p \alpha_A,
\end{equation}
where the usual ``case A'' recombination coefficient is \citep[table 2.1 in ][]{osterbrock_ferland_2006}, 
\begin{equation}
    \alpha_A = \sum_{i=1}^\infty \alpha_i \approx 4.18\times10^{-13}{\rm cm^3\,s^{-1}}\left(\frac{T_e}{10^4\,{\rm K}}\right)^{-0.72}
\label{eq:alphaA}
\end{equation}
While recombinations directly to the ground state replace a destroyed Lyman continuum photon, recombinations to $i>1$ do not (for the most part; if a recombination to $i>1$ occurs with a fast enough electron, it will create a photon above the Lyman edge, but this is a small effect). So, $i>1$ recombinations require a net depletion of the Lyman continuum flux. This sets a limit on how much of the wind the source Lyman continuum flux can photoionize: once $\int n_e n_p \alpha_B dV$ photons are consumed. Here  \citep[table 2.1 in ][]{osterbrock_ferland_2006}, 
\begin{equation}
\alpha_B = \sum_{i=2}^\infty \alpha_i\approx 2.59\times10^{-13}{\rm cm^3\,s^{-1}}\left(\frac{T_e}{10^4\,{\rm K}}\right)^{-0.83}
\label{eq:alphaB}
\end{equation} is the usual ``case B'' recombination coefficient, photoionization equilibrium collapses, and the wind recombines (i.e., Eq.~\ref{eq:electron_density_equilibrium} requires $n_e=0$ if $n_1\Gamma_1=\alpha_1n_en_p$ and all other $\Gamma_i=0$). 

In Fig.~\ref{fig:gamma_lyc_resid}, we show (for the marginally unbound winds) the ``net'' Lyman photoionization rate, $n_1\Gamma_1-\alpha_1 n_e n_p$, normalized to $\alpha_A n_en_p$. Here, we calculate $\alpha_1$ by subtracting Eq.~\ref{eq:alphaB} from Eq:~\ref{eq:alphaA}. There is net depletion of the Lyman continuum flux where the net rate is positive. Here, the dashed line shows the limiting photoionization rate that holds if all recombined atoms cascade to the ground state before being photoionized. Figs.~\ref{fig:gamma_lyc_resid}a-b show that at lower outflow rates, the Lyman continuum photoionizes at (nearly) the limiting rate before the Lyman transition radius. This indicates that the Lyman continuum is responsible for keeping the gas at the base of the wind photoionized. This is supported by Figs.~\ref{fig:optical_depth}a-b, which show that the effective Balmer (and Paschen) continuum optical depth is $\lesssim1$ here. This changes beyond the Lyman transition radius, because there the Lyman continuum flux is depleted, leaving only the Lyman photons locally produced by recombinations directly to the ground state. Indeed, in each run past the Lyman transition radius, the net Lyman photoionization rate is \footnote{We have not investigated the net negative residual here in detail, but expect it is likely because the tail of the recombination continua to higher hydrogen energy levels will inject some photons into the Lyman band. In any case, the residual is too small to affect our conclusions.} $\gtrsim -0.1$ (meaning $n_1\Gamma_1\approx\alpha_1 n_e n_p$). This indicates that the Lyman photons cannot photoionize the recombinations to $i>1$. If Lyman photons were the only photons capable of photoionizing the wind, the gas would then recombine; yet, it remains ionized.

The wind does not recombine past the Lyman radius because it is also photoionized by the Balmer continuum (and, to a lesser extent, Paschen and higher level continua). Before analyzing its role, it is important to clarify that $n=2$ acts as an ``effective'' ground state because of Lyman~$\alpha$ trapping. Any hydrogen atom excited to the $i=2$ state will soon cascade to $i=1$ and release a Lyman~$\alpha$ photon. If the Lyman~$\alpha$ photon escapes, the cascade to $i=1$ is irreversible. However, the optical depth to Lyman~$\alpha$ is usually orders of magnitude above unity in our winds. To see this, first note that the Lyman bound-free optical depth is extremely large past the Lyman radius (Fig.~\ref{fig:optical_depth}). Then, we can compare the Lyman~$\alpha$ Sobolev optical depth (Eq.~\ref{eq:tau_s}) with the local bound-free optical depth at the Lyman edge ($\tau_{\rm LyC}^{\rm (local)}=n_1\sigma_1 r$), 
\begin{equation}
    \tau_{\rm s}^{\rm (Ly\alpha)}/\tau_{\rm LyC}^{\rm (local)} \sim 4\times10^2\left(\frac{r}{10^{17}\,{\rm cm}}\right)^{1/2},
\end{equation}
where the Lyman~$\alpha$ optical depth is strictly larger. Here, we assumed $n_2\ll n_1$, took $|\frac{dv}{ds}|=v/r$ and used the marginally unbound velocity profile (Eq.~\ref{eq:v_mu}). So, the vast majority of spontaneous $i=2\rightarrow1$ Lyman~$\alpha$ emissions are compensated for by $i=1\rightarrow2$ Lyman~$\alpha$ absorptions. This prevents the $i=2$ hydrogen population from collapsing, which facilitates Balmer photoionization. Later, in Sec.~\ref{sec:caveats}, we highlight a couple of important caveats related to the numerical treatment of Lyman~$\alpha$ trapping in \textsc{Sirocco}.

Now, return to Eq.~\ref{eq:electron_density_equilibrium}, and take $n_1\Gamma_1=\alpha_1n_en_p$ so that the Lyman continuum does not contribute to net photoionization. If $\Gamma_i=0$ for $i>2$, then the wind remains ionized only if $n_2\Gamma_2=\sum_{i=2}^\infty\alpha_i n_e n_p=\alpha_B n_e n_p$. Analogously to the Lyman continuum case we described already, this requires net depletion of the Balmer continuum flux, because only recombinations directly to $i=2$ repopulate the Balmer continuum\footnote{This statement neglects that recombinations to $i>2$ with fast enough electrons will produce photons with energies above $3.4$ eV. Although this approximation is not as strong as it is for the Lyman continuum, the effect is still not important enough to affect our conclusions.}. In Fig.~\ref{fig:gamma_bac_resid}, we show the net Balmer photoionization rate ($n_2\Gamma_2-\alpha_2 n_e n_p$), again normalized to $\alpha_A n_e n_p$. Here, we use the expression \citep[table 2.1 in ][]{osterbrock_ferland_2006}
\begin{equation}
    \alpha_2\approx 7.69\times10^{-14}{\rm cm^3\,s^{-1}}\left(\frac{T_e}{10^4\,{\rm K}}\right)^{-0.64}.
\label{eq:alpha2}
\end{equation}
In panels a-b, at radii within the Lyman transition radius, the net rate is negative (and order-unity), indicating that there is net \textit{production} of Balmer continuum photons. Indeed, Figs.~\ref{fig:optical_depth}a-b show that this inner region is optically thin to Balmer photoionization and Fig.~\ref{fig:gamma_lyc_resid}a-b shows that the Lyman photons photoionize nearly every recombined atom. This leads to the degradation of the Lyman photon into multiple lower energy photons.  For example, consider a recombination to $i=2$ that then cascades to $i=1$ before being photoionized. This produces a Balmer continuum photon and a Lyman~$\alpha$ photon (which is added to the Balmer band) but costs a Lyman continuum photon. Thus, in this inner region, the depletion of the Lyman continuum flux adds to the Balmer continuum flux, leading to the negative residual. Beyond the Lyman transition, the story changes. The Lyman continuum flux no longer contributes to net photoionization, and the net Balmer photoionization rate becomes positive. This indicates that now the Balmer continuum flux is responsible for net photoionization of the wind and is thus being depleted. In these regions, $n_2\Gamma_2\approx \alpha_B n_e n_p$, indicating that the Balmer photons are responsible for photoionizing most of the atoms that recombined to energy levels above $i=1$. 

At higher outflow rates, the results are slightly different. In Figs.~\ref{fig:gamma_lyc_resid}c-d, even before the Lyman transition radius, the net Lyman photoionization rate is below its limiting value. This indicates that some of the recombined atoms at energy levels $i>1$ are photoionized before they cascade to $i=1$. Still, beyond the Lyman transition radius, $n_1\Gamma_1\approx \alpha_1 n_e n_p$, as was found for the lower outflow rates. In Figs.~\ref{fig:gamma_bac_resid}c-d, the net Balmer photoionization rate again increases beyond the Lyman radius, but does not reach its limiting value. This requires that recombined atoms at energy levels $i>2$ are photoionized before they cascade. Indeed, Figs.~\ref{fig:optical_depth}c-d show that these two runs are marginally opaque to Paschen photoionization. 

In Fig.~\ref{fig:gamma_pac_resid}, we show the net Paschen photoionization rate as well, where we use the $i=3$ recombination rate \citep[table 2.1 in ][]{osterbrock_ferland_2006},
\begin{equation}
    \alpha_3\approx 4.55\times10^{-14}{\rm cm^3\,s^{-1}}\left(\frac{T_e}{10^4\,{\rm K}}\right)^{-0.75}.
\label{eq:alpha3}
\end{equation}
The net rate is usually negative, indicating a net production of Paschen photons. Analogous to the net production of Balmer photons occurring before the Lyman transition radius in Fig.~\ref{fig:gamma_bac_resid}a-b, this is due to photon degradation. For instance, if a recombination to $i=3$ cascades to $i=2$ before being photoionized by the Balmer continuum, then one Balmer photon was spent to produce two Paschen photons. This explains why Paschen jumps are observed in some of the spectra in Fig.~\ref{fig:spectra} \citep[Paschen jumps are also present in some observed LRDs,][]{sneppen_2026_paschen}.

In Figs.~\ref{fig:gamma_pac_resid}c-d, the net Paschen photoionization rate is closer to zero, indicating that most of the $i=3$ recombined atoms are photoionized by the Paschen continuum before they cascade to lower levels. This explains why in Fig~\ref{fig:gamma_bac_resid}d the net Balmer photoionization rate is below its limiting case in these regions: Paschen photoionization partially intercepts the cascade to $i=2$. We caveat that this is affected by H$\alpha$ trapping, which may be affected by the Sobolev approximation (see Sec.~\ref{sec:caveats}) and should be explored further in a later work. Near the scattering surface in Fig.~\ref{fig:gamma_pac_resid}, the net Paschen photoionization is positive and order-unity. This is coincident with a decrease in the net Balmer photoionization at the scattering surface in Fig.~\ref{fig:gamma_bac_resid}. Note, an inset panel of $J_\nu$ is shown at the scattering surface for this run in Fig.~\ref{fig:optical_depth}d; this shows an extremely strong Balmer break, indicating that the Balmer continuum flux is depleted at this point. Yet, the wind remains ionized for an additional $\gtrsim10r_{\rm sph}$, suggesting that Paschen photoionization is pushing the recombination front out slightly farther. Still, these runs do not exhibit a Paschen break, indicating that this is not a dominant effect. Still, it is likely that at even higher outflow rates, detectable Paschen breaks do form. These would represent an extreme subclass of Paschen break LRDs that have not yet been discovered. 

Given our findings, we suggest that the outer part of the wind be called the ``Balmer cocoon''. The defining feature of the cocoon is that it is kept ionized by net Balmer photoionization (rather than net Lyman photoionization, which is typically the case). As discussed, this comes at the expense of the Balmer continuum flux, which is continuously depleted in the cocoon. Whether or not the Balmer continuum is significantly depleted depends on both the shape and normalization of the wind's density profile. At lower outflow rates, the wind forms either no Balmer break or a weak one (first two columns of Fig.~\ref{fig:spectra}). The Balmer cocoon will also end, by definition, if the entire Balmer continuum flux is depleted. The wind will then recombine (or, as in Fig.~\ref{fig:gamma_pac_resid}, may be kept ionized if there is net photoionization at higher energy levels). This region deserves to be studied in more detail in future work, as it appears responsible for many of the observational features of LRDs. 

There are several properties of the wind that facilitate the formation of the Balmer cocoon (but are not essential to its definition). The wind is optically thick to electron scattering, which boosts the effective optical depth to photoionization. Lyman~$\alpha$ is highly trapped, which boosts $n_2$ and thus the optical depth to Balmer photoionization. The source spectrum that we input at $r_{\rm sph}$ is cool, ranging from $T_{\rm bb}\approx1.4-4.9\times10^4\,{\rm K}$. At $T_{\rm bb}=4.9\times10^4\,{\rm K}$ (runs \texttt{mu\_m2.5} and \texttt{pt\_m2.5}), the ratio of photons between $3.4$ and $13.6$ eV to those above $13.6$ eV is $\approx2$; the Balmer photons barely dominate in population. At $T_{\rm bb}=1.4\times10^4\,{\rm K}$ (run \texttt{pt\_m15}), the ratio is $492$, and the Lyman continuum flux is negligible. This dramatic change is because the Lyman continuum is in Wien's tail of the blackbody distribution, so it is exponentially sensitive to the temperature.

% Still, accurately modeling Lyman~$\alpha$ escape requires tracking frequency redistribution, whereas \textsc{Sirocco} adopts the simplified Sobolev approximation (see Sec.~\ref{sec:caveats}). These details are unimportant for $n_2$ in the limit that the Balmer photoionization time is shorter than the Lyman~$\alpha$ escape time, because then an atom in $n=2$ is far more likely to be photoionized than to lose its Lyman~$\alpha$ photon to frequency or spatial diffusion. In this limit, Balmer photoionization (not Lyman~$\alpha$ escape) regulates the $n=2$ population, and Eq.~\ref{eq:n2} holds independent of the transfer details. The enormous $\tau_{{\rm Ly}\alpha}$ favors this limit, but future work should verify where it holds. 

 %% FIGURE N2 FULL PAGE
\begin{figure*}
\centering
\includegraphics[width=\textwidth]{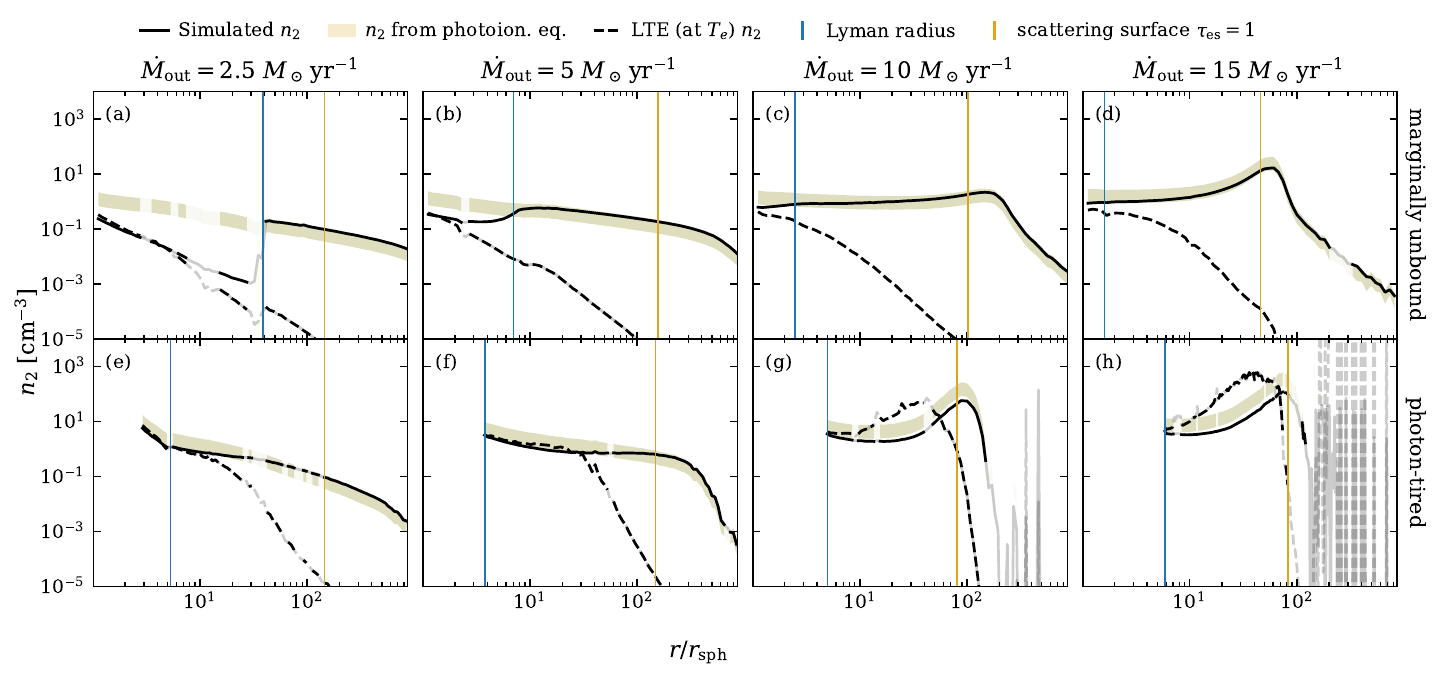}
\caption{Number density of neutral Hydrogen in the $n=2$ state, $n_2$, in the marginally unbound winds (top row) and the photon-tired winds (bottom row). We plot the simulated $n_2$ (black solid lines), the LTE $n_2$ value at the local electron temperature (black dashed lines), and the predicted range of $n_2$ given photoionization equilibrium with the Balmer continuum (green shaded region; the bounds of the region are $\alpha_2 n_e n_p/\Gamma_2<n_2<\alpha_B n_e n_p/\Gamma_2$, see Sec.~\ref{sec:photoionization}). We also show the location of the Lyman radius (inside-out $\tau_{\rm LyC}=3$, beyond which Lyman photons do not contribute to net photoionization of the gas). In all cases, beyond the Lyman radius, $n_2$ is accurately predicted by photoionization equilibrium by the Balmer continuum -- this is the ``Balmer cocoon''.}
\label{fig:n2}
\end{figure*}

%%%% DISCUSSION %%%%
\section{Discussion}
\label{sec:disc}

\subsection{Plausible accretion rates in LRDs} 
\label{sec:plausible_accretion_rates}

In this work, we produced LRD-like SEDs with mass accretion rates $\sim5-15\,M_\odot\,{\rm yr}^{-1}$. Is this physically reasonable? LRDs are preferentially hosted by $\gtrsim10^{11}\,M_\odot$ dark matter halos \citep{arita_2025,pizzati_2025,lin_2026}, which are $\sim1-2$ orders of magnitude lighter than the halos that host UV-bright quasars \citep{porciani_2004,croom_2005}. The lower-mass dark matter halos that host LRDs may sustain baryonic inflow rates $\gtrsim100\,M_\odot\,{\rm yr}^{-1}$ at $z\sim 5$ \citep{claudeandre_2011,dekele_2017,inayoshi_2026}. Only a fraction of this gas feeds the SMBH. Unless the SMBH fueling is extremely bursty, it is difficult to achieve accretion rates comparable to the halo inflow rate. Furthermore, more and more LRDs are being detected at lower redshift \citep{lin_2026_lrd,lin_2026_lrdsurvey,ji_2026_lordoflrds,ding_2026}, where there is less available gas and the fueling constraints are stricter. In our work, complete Balmer breaks form at $\dot{M}_{\rm out}\gtrsim 10-15\,M_\odot\,{\rm yr}^{-1}$, and the characteristic V-shape forms at $\dot{M}_{\rm out}\gtrsim 5\,M_\odot\,{\rm yr}^{-1}$. While large, these accretion rates are plausible because they are well within the halo inflow rate.

Many LRD models focus on the required column density to produce a Balmer break, but not on the outflow rate this implies. Given that we know the characteristic velocity scale ($\sim 300\,{\rm km\,s^{-1}}$) and length scale ($\sim10^{17}\,{\rm cm}$) of LRD envelopes, we can use the outflow rate (which is of order the fueling rate, see Sec.~\ref{sec:outflow_problem}) to solve for the photospheric density via mass conservation. Here, we assume that the photosphere is part of the wind -- rather than being set by a hydrostatic structure or dense inflow that lies underneath an optically thin wind. Then, given a maximum fueling rate, the photospheric density has an upper bound, 
\begin{equation}
\begin{aligned}
    \rho_{\rm LRD} &\lesssim 10^{-15}\,{\rm g\,cm^{-3}}\\&\times\left(\frac{\dot{M}_{\rm LRD,max}}{100\,M_\odot\,{\rm yr}^{-1}}\right)\left(\frac{r_{\rm LRD}}{10^{17}\,{\rm cm}}\right)^{-2}\left(\frac{v_{\rm LRD}}{300\,{\rm km\,s^{-1}}}\right)^{-1}
    \label{eq:max_rho_LRD}
\end{aligned}
\end{equation}

Yet, LTE models of LRDs require densities $\sim10^{-13}\,{\rm g\,cm^{-3}}$ \citep{kido_2025,liu_2025,begelman_dexter_2026,zhou_2026} to produce a Balmer break, two orders of magnitude above this rough limit. However, the outer parts of our winds (which we describe as ``Balmer cocoons'', see Sec.~\ref{sec:photoionization}) are kept ionized by Balmer photons. This causes the $n=2$ density to far exceed its LTE value, allowing us to produce the phenomenology of LRDs at much lower densities and outflow rates. We show this in Fig.~\ref{fig:n2}, where we compare the $n_2$ density (solid black line) to the LTE value (black dashed line) as a function of radius for each \textsc{Sirocco} calculation. The shaded regions bound the two limiting cases for Balmer photoionization equilibrium ($\alpha_2 n_e n_p/\Gamma_2<n_2<\alpha_B n_e n_p/\Gamma_2$). Figs.~\ref{fig:n2}g-h instead show that, sometimes, the LTE $n_2$ prediction exceeds the simulated $n_2$; this is because the electron temperatures here are extremely low (see Fig.~\ref{fig:temp_ion}). 

Interestingly, NLTE \textsc{Cloudy} models also require denser, $\sim10^{-13}-10^{-15}\,{\rm g\,cm^{-3}}$ gas \citep[where the $n_2$ population may be sustained by collisions,][]{inayoshi_2025}. However, \textsc{Cloudy} does not accurately model electron scattering at high optical depths, which plays an important role in boosting the effective optical depth to Balmer photoionization (Fig.~\ref{fig:optical_depth}).  Although Eq.~\ref{eq:max_rho_LRD} is crude, we expect that the broad lesson is robust: Balmer cocoons produce breaks at much lower densities than LTE (or, to a lesser degree, NLTE models absent electron scattering) envelopes, allowing $\dot{M}_{\rm out}$ to be lower.

\subsection{Supermassive and quasi- stars}

We have argued that LRDs are powered by super-Eddington accretion onto a SMBH. Instead, LRDs might be ``supermassive stars'' \citep{nandal_loeb_2026} or ``quasi-stars'' \citep{begelman_dexter_2026}. Supermassive stars are hypothetical stellar structures that may form from pristine, inefficiently cooling gas in the early universe \citep{loeb_rasio_1994,bromm_loeb_2003} or by runaway stellar mergers \citep[even if the gas is metal-enriched,][]{nandal_chon_2026}. Supermassive stars ultimately collapse to a SMBH \citep{baumgarte_shapiro_1999}. Quasi-stars are similar, except the core \textit{is} the SMBH; quasi-stars may form in the aftermath of the collapse of a supermassive star. In either case, the slow winds we advocated for in Sec.~\ref{sec:fountain_flow}-\ref{sec:photon_tired_winds} are equally applicable; then, the ``spherization radius'' (Eq.~\ref{eq:spherization_radius}) of the disk is instead the surface of the star. 

Although supermassive and quasi- stars are both attractive candidates for LRDs, their basic nature remains unclear.  Various works have found that supermassive stars may form given the right environments \citep[e.g.][and references therein]{latif_2013,chon_2020,reinoso_2023}, but we do not know if these environments are common enough or if the star survives long enough to explain LRDs. In quasi-stars, convection must vigorously transport the  energy produced by accretion outwards \citep[e.g.,][]{eliot_2000}, preventing too much energy from being produced near the horizon. In simulations, radiatively-inefficient accretion flows instead develop biconical inflow-outflow structures, and the energy production occurs near-horizon (see Sec.~\ref{sec:fast_agn_winds}). This would likely unbind a quasi-star. However, quasi-stars have not been simulated \citep[although, see the \textsc{MESA} calculations by][]{hassan_2026} and their stability remains unknown.

\subsection{SS433: a stellar-mass analogue of LRDs?}
\label{sec:ss433}

Are there stellar-mass analogues of LRDs? The best case, as suggested by \cite{zhou_2026}, is the microquasar SS433 \citep[for a review, see][]{fabrika_2004}. In SS433, the compact object accretes gas from an orbiting donor star at a rate $10^2-10^3$ times above the Eddington limit. We view SS433 nearly edge-on. The entire disk is obscured by a wind photosphere that extends to $\gtrsim10^{12}\,{\rm cm}$. This is roughly $10^2$ times larger than its spherization radius, and $10^5-10^6$ times larger than the gravitational radius. Notably, the entire structure precesses on about a half-year period \citep{fabian_rees_1979,abell_margon_1979}, allowing observations to probe the lateral dependence of its photosphere. Such observations have revealed ``equatorial'' winds with velocities ranging from $\gtrsim1500\,{\rm km\,s^{-1}}$ at higher latitudes to $\sim100\,{\rm km\,s^{-1}}$ near the equator \citep{fabrika_1997}. Furthermore, the spectrum of SS433 is qualitatively similar to that of a Wolf-Rayet (WR) star \citep{vandenheuvel_1981,fuchs_2006}, wherein the stellar surface is cloaked by a dense wind. This inflated, slow-moving, star-like wind photosphere is remarkably similar to the picture we advocated for in Sec.~\ref{sec:lrds_and_stars}.

 It is exciting that SS433, like LRDs, features slow winds, but it only does so at shallow viewing angles. If LRDs had the same velocity structure, then their covering factor could not be too large, or else their winds would often be faster than observed. Indeed, many authors have argued that LRD envelopes cover $\sim50\%-100\%$ of the central engine \citep[e.g.,][]{inayoshi_2025,yan_2026,lin_2026_lrd,torralba_2026}, which would indicate that LRDs do not share the same lateral structure as SS433. On the other hand, the detection of high-ionization narrow lines in some LRDs suggests there must sometimes be undetected, anisotropic ultraviolet emission, possibly from a polar funnel \citep{tang_2026,ji_2026,sok_2026}. \cite{madau_maiolino_2026} alternatively suggested that LRDs and LBDs belong to the same population and are distinguished by viewing angle; then, LRDs would have a covering factor of $\sim20\%$ (however, if LRDs and LBDs emerge from the same population of super-Eddington accretors, we expect that the outflow rate is degenerate with the viewing angle).

Whatever their covering fraction, LRD winds may not have the same acceleration mechanism as in SS433, and thus have a different velocity structure. For instance, the winds of WR stars are likely accelerated by the iron opacity bump, which is triggered where the envelope temperature crosses $\approx1.8\times10^5\,{\rm K}$ \citep{ro_matzner_2016}. The same may be true of SS433's winds, which appear WR-like \citep{vandenheuvel_1981,fuchs_2006}. Conversely, LRD envelopes are too cool to tap the iron opacity bump, and low metallicities further decrease its importance. The Helium opacity bump may play a role, but it is far weaker than both the iron opacity bump and the electron scattering opacity, especially at LRD densities. This is why we instead invoke scattering-driven, photon-tired winds in Sec.~\ref{sec:photon_tired_winds}. If LRD winds are poorly accelerated, then they are slower and more pressurized at the same outflow rate, which may cause the covering factor of the slow winds in LRDs to be higher than that of SS433.

\subsection{Comparison to \cite{sneppen_2026a} models}

\cite{sneppen_2026a,sneppen_2026_paschen,sneppen_2026_lbd} used \textsc{Sirocco} to show that AGN shrouded in dense winds may explain LRDs. Our radiative transfer calculations are similar to theirs. They use both $\beta$-law and constant-velocity winds, while we use ``marginally unbound'' (Sec.~\ref{sec:fountain_flow}) and ``photon-tired'' (Sec.~\ref{sec:photon_tired_winds}) winds. The resulting SEDs do not significantly differ; our intent is to explain \textit{why} such slow, dense winds are dynamically feasible in the first place (Sec.~\ref{sec:outflow_problem}-\ref{sec:lrds_and_stars}). \cite{sneppen_2026a} pointed out that emission line photons must scatter through both inflowing and outflowing gas to explain the symmetry of their exponential wings. They attributed the inflowing component to a thick torus; we expect the disk to instead be thin and obscured by a marginally unbound wind, which manifests as a fountain flow wherein some streamlines fall back (Sec.~\ref{sec:fountain_flow}). They attribute the narrow emission line cores to Keplerian rotation, whereas here we do not include rotation -- the narrow cores emerge naturally from the photosphere, and their width is set by the photospheric velocity (Fig.~\ref{fig:halpha}). We favor this interpretation, because each wind streamline carries the specific angular momentum from where it was launched, which is negligible compared to the Keplerian angular momentum at the photosphere. Furthermore, we analyzed the wind's photoionization structure (Sec.~\ref{sec:photoionization}), which we interpret as a ``Balmer cocoon'' --  a scattering-thick region kept ionized by the depletion of Balmer continuum photons rather than Lyman continuum photons. Importantly, the NLTE nature of the region allows Balmer breaks to be formed at much lower accretion rates than in LTE models (Sec.~\ref{sec:plausible_accretion_rates}). 

\subsection{Caveats and Future Work}
\label{sec:caveats}
\textit{Sobolev approximation.} \textsc{Sirocco} treats bound--bound absorption in the Sobolev approximation \citep{sobolev_1957,CAK75}, which exploits the large velocity gradient, $\mathrm{d}v/\mathrm{d}s$, in a supersonic outflow. As a photon propagates through the accelerating wind, the bulk velocity along its path changes, continuously Doppler-shifting its frequency in the co-moving frame of the gas. A given transition is resonant only over the Sobolev length, $\ell_S \simeq \Delta v/(\mathrm{d}v/\mathrm{d}s)$ -- the distance over which the co-moving frequency stays within one line width, $\Delta v$, of line center. Usually, $\Delta v$ is the thermal line width, which is $\sim10\,{\rm km\,s^{-1}}$. Since the wind is supersonic, $\ell_S\ll r$ throughout. However, this
assumption breaks down for lines where $\Delta v\gg v_{\rm th}$ -- namely, Lyman~$\alpha$, where the line profile has large damping wings such that $\Delta v\sim10^2-10^3\,{\rm km\,s^{-1}}$ \citep{dijkstra_2019}. Thus, Lyman~$\alpha$ requires a full resonant scattering treatment with frequency redistribution \citep{neufeld_1990}. In our work, the $n=2$ population is sustained by Lyman~$\alpha$ trapping, so Balmer photoionization may be sensitive to these details. Still, given that in either the Sobolev approximation or an improved resonant scattering treatment Lyman~$\alpha$ is strongly trapped, we expect that the $n=2$ population will remain large enough to be opaque to the Balmer continuum.  Quantitatively establishing this with more detailed microphysics would be a valuable direction for future work.  It will also be valuable to better understand the structure of the Balmer cocoon and the recombination front, e.g., can the hydrogen level populations be determined analytically. Since we have only simulated a small fraction of the possible parameter space, it would be particularly useful to determine the properties and location of the recombination front semi-analytically as a function of the mass-loss rate, the base luminosity, density profile, and other parameters. This region strongly affects the observed continuum and lines.  
%The exact density profile may be important because the marginal unbound winds with $n \propto r^{-3/2}$ have a roughly equal number of recombinations per decade in radius while for a constant speed outflow with $n \propto r^{-2}$ the total number of recombinations at large radii declines $\propto r^{-1}.$   The ionization structure of the latter may be more subtle as a result.}

Interestingly, resonant scattering may also be important for Balmer photons \citep[e.g.,][]{chang_2026}, which regulates the $n=3$ population. We find that the highest density winds are marginally opaque to the Paschen continuum (Fig.~\ref{fig:optical_depth}) and have signatures of Paschen bound-free absorption (Fig.~\ref{fig:spectra}h), which sensitively depend on H$\alpha$ trapping. Indeed, some LRDs feature double-peaked Balmer lines, thought to arise from frequency diffusion into the less-opaque wings of the line profile \citep{naidu_2025}. 
 
\textit{Accretion Flow Structure.} We assumed a thin, radiatively-efficient outer disk that transitions at $r_{\rm sph}$ to a quasi-spherical, radiation-pressure-supported inner torus, from which a slow, off-axis, ``photon-tired'' wind is launched -- requiring that the luminosity budget be barely large enough to accelerate the outflow. In our simplified one-dimensional model, we parameterized both the outflow rate and luminosity, and assumed that the gas sonic point lies within $r_{\rm sph}$. Although the resulting picture is consistent with observations, which clearly require slow outflows, the acceleration of the wind and the transition region itself remain poorly  understood.

For instance, our one-dimensional model is laminar, yet radiation-pressure-driven winds in massive stars are unstable and develop underdense, porous channels through which radiation escapes \citep[e.g.,][]{owocki_2004}. On average this may lower the ``effective'' Eddington ratio below unity. Then, acceleration may not occur until the photosphere, where porosity no longer suppresses the effective Eddington ratio.

In a realistically biconical wind, the steady-state structure is set by lateral force balance. Rotation is then important, since the centrifugal force decollimates the wind whereas the hoop stress of a wound, magnetocentrifugal wind would collimate it more strongly. Capturing the true two-dimensional structure requires accounting for both. However, strong magnetocentrifugal forces accelerate winds easily, so it is unlikely that they power the slow LRD winds. Notably, many LRDs show high-ionization lines \citep{tang_2025,tripodi_2025b,ji_2026}, which may be produced by the fast, on-axis wind and scattered into our line of sight, emphasizing the need to understand the biconical nature of LRD winds.

Finally, if the wind is truly a fountain flow, it is likely time-dependent: streamlines fail, collapse and relaunch (albeit on a dynamical timescale, which is $\sim10^2\,{\rm yr}$ at the $\sim10^{17}\,{\rm cm}$ photosphere, and further slowed by cosmological time dilation). At much larger radii, the disk may also become unstable to star formation, which we have neglected. Fully capturing these effects requires multiscale numerical simulations of super-Eddington AGN. 

\subsection{Summary}
\label{sec:disc:summary}

Figure~\ref{fig:cartoon} is an infographic that summarizes the salient features of our model, which we list here:

\begin{itemize}
\item \textbf{The slow outflows of LRDs pose an energy-budget problem for super-Eddington accretion models that requires an off-axis line of sight (Sec.~\ref{sec:outflow_problem}, Fig.~\ref{fig:cartoon}-\ref{fig:velocities}).} LRDs show slow, $\sim100-300\,{\rm km\,s^{-1}}$ outflows, yet a super-Eddington
torus must shed most of its viscously generated energy into far faster winds, $\gtrsim10^3-10^4\,{\rm km\,s^{-1}}$. Since we observe slow gas, the fast wind must escape near the
poles where it is collimated into funnels, while we view the system off-axis.

\item \textbf{Radiation pressure launches a slow, marginally unbound outflow from the spherization radius, $r_{\rm sph}$ (Sec.~\ref{sec:fountain_flow}-\ref{sec:photon_tired_winds}).} Beyond $r_{\rm sph}$ the disk is thin and cools efficiently;
within it, radiation pressure inflates the disk into a quasi-spherical envelope. The outflow may be ``photon-tired'' \citep{owocki_1997}, meaning that the radiation can only barely unbind the gas. This happens when the radiation field is only marginally super-Eddington -- a condition met precisely at $r_{\rm sph}$. Photon-tired winds are marginally unbound, such that some streamlines escape while others fall back. This forms a fountain flow, whose mix of inflowing and outflowing gas is necessary to symmetrize the broad, exponential (electron-scattering) wings of the line profiles. The same type of fountain flows may be launched by supermassive stars or quasi-stars, in which case our black hole accretion inspired models here would be applicable to a broader class of LRD central engines.

\item \textbf{The Balmer break scales with the outflow rate (Sec.~\ref{sec:spectra}, Fig.~\ref{fig:spectra}).} At $\dot{M}_{\rm out}\sim2.5\,M_\odot\,{\rm yr}^{-1}$ the SED is breakless, resembling a ``little blue dot''
\citep{brazzini_2026}. At $\sim5-10\,M_\odot\,{\rm yr}^{-1}$ a partial break lets some blue light escape, giving the ``V-shaped'' SED seen in many LRDs \citep{setton_2025_vshape}. At
$\sim15\,M_\odot\,{\rm yr}^{-1}$ the break is complete and erases the blue emission entirely, as in the reddest LRDs \citep[e.g.,][]{degraaff_2025,naidu_2025}. We expect this trend with $\dot{M}_{\rm out}$ to be robust, though the exact boundaries are model-dependent and subject to a viewing-angle dependence we have not modeled. This $\dot{M}_{\rm out}$ sequence mirrors the column-density sequence found by \cite{sneppen_2026a}.

\item \textbf{Balmer emission lines with P~Cygni profiles atop broad, electron-scattering wings naturally emerge from our wind models (Sec.~\ref{sec:spectra}, Fig.~\ref{fig:halpha}).} Each line has a narrow emission core, where the width is near the photospheric velocity. Beyond the photosphere, the gas recombines, usually leading to a blueshifted absorption trough that also traces the velocity of the recombined gas. The width of the absorption trough is broader in the marginally unbound winds and narrower in the photon-tired winds; this is because the former have a velocity gradient in the absorbing medium, while the latter do not. This suggests that the shape of the absorption trough in LRDs encodes more complex information about the velocity of the absorbing medium. The emission lines in our models also feature broad ($\gtrsim10^3\,{\rm km\,s^{-1}}$) exponential wings that result from scattering in the dense photospheric gas, rather than bulk motion.

\item \textbf{Most of the wind is photoionized by the Balmer continuum (Sec.~\ref{sec:photoionization}, Fig.~\ref{fig:gamma_lyc_resid}-\ref{fig:n2}).} Ionizing radiation usually means photons that ionize hydrogen from the ground state ($h\nu>13.6\,{\rm eV}$); we
find that these are quickly consumed and degraded to longer wavelengths. As the gas then tries to recombine, its $n=2$ population -- which is sustained by Lyman~$\alpha$ trapping, making $n=2$ an effective ground
state -- grows opaque to the Balmer continuum ($h\nu>3.4\,{\rm eV}$). The Balmer continuum keeps the wind ionized out to the electron-scattering photosphere (at
$\sim10^{17}\,{\rm cm}$) or wherever it is depleted, beyond which the gas recombines. This gas is firmly out of LTE; the heating (dominated by photoionization) and cooling (dominated by free-free and, usually, lines) processes are not in detailed balance, and $n=2$ is boosted orders of magnitude above its LTE value. We suggest ``Balmer cocoon'' as a name for this region. It plays an important role in shaping the observational appearance of the LRD, so it should be studied further in future work. 

\item \textbf{The Balmer-photoionized, scattering-thick wind lets Balmer breaks form at lower densities (Sec.~\ref{sec:plausible_accretion_rates}),} given that the photosphere is part of the wind. In LTE models
\citep[or, possibly, in NLTE models with simplified treatments of electron scattering,][]{inayoshi_2025}, $n=2$ is populated collisionally, requiring densities $\gtrsim10^{-13}\,{\rm g\,cm^{-3}}$ \citep[][]{liu_2025} and thus outflow rates $\gtrsim10^4\,M_\odot\,{\rm yr}^{-1}$
(Eq.~\ref{eq:max_rho_LRD}), far above the inflow rates of the lower-mass halos that host LRDs \citep{claudeandre_2011,dekele_2017,arita_2025}. Because our $n=2$ population can far exceed its LTE value, and because electron scattering further boosts the effective optical depth to the Balmer continuum, we match the same phenomenology at much lower (and more plausible) rates,
$\sim5-15\,M_\odot\,{\rm yr}^{-1}$. Although the details are model-dependent, Balmer cocoons generically lower the density (and thus $\dot{M}_{\rm out}$) required to produce LRD-like spectra. 
\end{itemize}

The analysis scripts used to make all figures (excluding Figs.~\ref{fig:cartoon}-\ref{fig:velocities}), along with the \textsc{Sirocco} run files and minor modifications to the \textsc{Sirocco} source code, are made publicly available \href{https://github.com/nkaaz/LRDs-as-Super-Eddington-Fountain-Flows_REPRODUCIBLE}{here}.

\vspace*{5pt}

\begin{acknowledgments}
We thank Jonatan Jacquemin-Ide, Rostom Mbarek, Mor Rozner, Bruce Draine, Jenny Greene, Bingjie Wang, David Setton and Enrico Ramirez-Ruiz for useful discussions and comments. NK is jointly funded by postdoctoral fellowships at the Princeton Center for Theoretical Sciences and the Princeton Gravity Initiative. Much of the coding was performed with assistance from Claude Opus 4.8. We also used Claude Opus 4.8 for simulated peer-review. 
\end{acknowledgments}

%%%%%%%%%%%%%%%
%% APPENDIX  %% 
%%%%%%%%%%%%%%%
\appendix

\section{Absorption troughs in LRD Spectra}
\label{app:obs_vel}

In Table~\ref{tab:lrd_absorption}, we list the velocity offsets of the absorption features detected in LRD spectra from various sources. Not all cited works explicitly listed the velocities or included error bars. In cases where the velocities were not listed, we measured the velocities by hand from the spectra. Each velocity is used in Fig.~\ref{fig:velocities}. 

% Requires \usepackage{booktabs} and a bibliography package (natbib/aastex) for \citet
\begin{table*}
\centering
\caption{Blueshifted and redshifted absorption velocities measured in little red dots (LRDs).
Velocities are relative to systemic; negative values indicate blueshifts (outflows),
positive values indicate redshifts (inflows).}
\label{tab:lrd_absorption}
\begin{tabular}{lccl}
\toprule
Source & Velocity (km\,s$^{-1}$) & Line & Reference \\
\midrule
GOODS-N-9771         & $-340$                 & H$\alpha$      & \citet{matthee_2024} \\
J1148-18404          & $+50$                  & H$\alpha$      & \citet{rusakov_2026} \\
\midrule
JADES-GN-68797       & $-272$                 & H$\alpha$      & \citet{rusakov_2026} \\
RUBIES-EGS-42046     & $-371$                 & H$\alpha$      & \citet{rusakov_2026} \\
RUBIES-EGS-49140     & $-82$                  & H$\alpha$      & \citet{rusakov_2026} \\
\midrule
CEERS 7902   & $-265$                 & H$\beta$       & \citet{kocevski_2025} \\
CEERS 10444   & $-182$                 & H$\alpha$      & \citet{kocevski_2025} \\
CEERS 13318   & $-401$                 & H$\alpha$      & \citet{kocevski_2025} \\
\midrule
RUBIES-BLAGN-1       & $-231^{+24}_{-20}$     & He\,\textsc{i} & \citet{wang_2025} \\
\midrule
GN-28074             & $-351^{+14}_{-16}$     & H$\alpha$      & \citet{juodzbalis_2024} \\
GN-28074             & $-351^{+14}_{-16}$     & H$\beta$       & \citet{juodzbalis_2024} \\
GN-28074             & $-506\pm7$             & He\,\textsc{i} & \citet{juodzbalis_2024} \\
\midrule
J0923P0402-BHAE-1    & $-218$                 & H$\alpha$      & \citet{lin_2024} \\
J1526M2050-BHAE-1    & $-194$                 & H$\alpha$      & \citet{lin_2024} \\
J1526M2050-BHAE-3    & $-67$                  & H$\alpha$      & \citet{lin_2024} \\
\midrule
JADES/BlackTHUNDER 159717 & $-26\pm4$         & H$\alpha$      & \citet{deugenio_2026a} \\
\midrule
Abell2744-QSO1       & $-40\pm10$             & H$\alpha$      & \citet{deugenio_2026b} \\
\midrule
The Cliff            & $+49^{+26}_{-21}$      & H$\alpha$      & \citet{degraaff_2025} \\
\midrule
MoM-BH*-1            & $+42^{+80}_{-200}$     & H$\beta$       & \citet{naidu_2025} \\
MoM-BH*-1            & $+1556^{+232}_{-1378}$ & H$\beta$       & \citet{naidu_2025} \\
MoM-BH*-1            & $-1532^{+345}_{-113}$  & H$\beta$       & \citet{naidu_2025} \\
\midrule
PAN-BH*-1            & $-94\pm4$              & H$\alpha$      & \citet{torralba_2026} \\
\midrule
J1025+1402           & $+73$                  & H$\beta$       & \citet{lin_2026_lrd} \\
J1025+1402           & $-134$                 & H$\alpha$      & \citet{lin_2026_lrd} \\
J1025+1402           & $-613$                 & He\,\textsc{i} & \citet{lin_2026_lrd} \\
J1025+1402           & $-410$                 & He\,\textsc{i} & \citet{lin_2026_lrd} \\
J1025+1402           & $-248$                 & He\,\textsc{i} & \citet{lin_2026_lrd} \\
J1022+0841           & $-95$                  & H$\alpha$      & \citet{lin_2026_lrd} \\
J1022+0841           & $+102$                 & H$\alpha$      & \citet{lin_2026_lrd} \\
J1022+0841           & $-55$                  & He\,\textsc{i} & \citet{lin_2026_lrd} \\
\midrule
GLIMPSE-17775        & $-200\pm32$            & H$\alpha$      & \citet{kokorev_2026} \\
GLIMPSE-17775        & $-52\pm83$             & He\,\textsc{i} & \citet{kokorev_2026} \\
GLIMPSE-17775        & $+45\pm22$             & He\,\textsc{i} & \citet{kokorev_2026} \\
\bottomrule
\end{tabular}
\end{table*}

\section{Comparison of Truncated Spectra}
\label{app:truncated_spectra_comparison}

As discussed in Sec.~\ref{sec:numerical}, runs \texttt{mu\_m15} and \texttt{pt\_m15} exhibit recombination fronts, beyond which the calculations are poorly converged. This happens because the high densities completely deplete the ionizing photon population, so there are no photons to heat the gas beyond the recombination front. Yet, the density in this region is high enough to artificially reprocess the spectra. For this reason, we performed analogous ``truncated'' calculations of these two runs, where we excised the gas beyond $8.5\times10^{18}\,{\rm cm}$ and $5.2\times10^{17}\,{\rm cm}$ for \texttt{mu\_m15} and \texttt{pt\_m15}, respectively.

\begin{figure}%[bht]
    \centering
    \includegraphics[width=\textwidth]{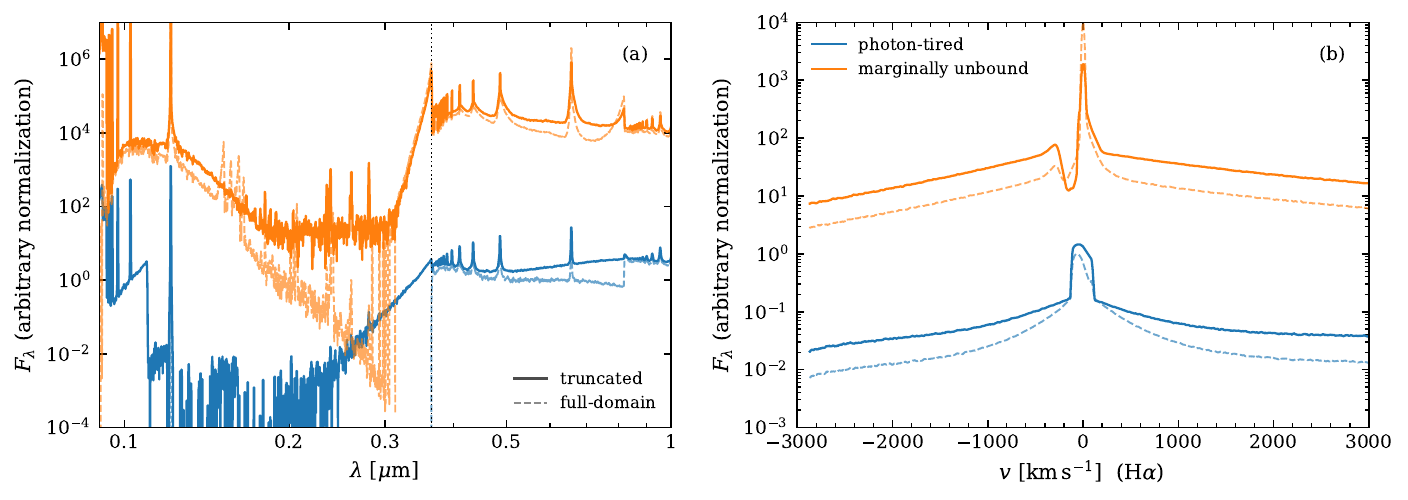}
    \caption{Comparison of the continuum (panel a) and H$\alpha$ (panel b) spectra in \texttt{mu\_m15} and \texttt{pt\_m15}. The scale on the $y$-axes is arbitrary. In the ``full-domain'' spectra, we retain the gas up to $r_{\rm out}=10^{20}\,{\rm cm}$ and $2\times10^{19}\,{\rm cm}$ for \texttt{mu\_m15} and \texttt{pt\_m15}, respectively. Here, the outer region is cold and poorly converged, partially reprocessing the spectra. In the ``truncated'' spectra, we performed the same calculations, but excised the unconverged gas beyond $8.5\times10^{18}\,{\rm cm}$ and $5.2\times10^{17}\,{\rm cm}$ for \texttt{mu\_m15} and \texttt{pt\_m15}, respectively. We also label the Balmer edge with a black dotted line.}
    \label{fig:truncated_spectra}
\end{figure}

We compare the continuum and H$\alpha$ spectra in Fig.~\ref{fig:truncated_spectra}. Reassuringly, the ``truncated'' and ``full-domain'' spectra (panel a, analogous to Fig.~\ref{fig:spectra})  are similar near the Balmer edge ($\lambda=0.36\,\mu{\rm m}$). In the truncated runs, more of the light blueward of the Balmer edge escapes. In the photon-tired run, this light is completely attenuated thanks to the large population of neutral hydrogen in the $n=2$ state (Fig.~\ref{fig:n2}). So, the truncated run underestimates the attenuation of blue light, while the full domain run attenuates it incorrectly. Thus, emission blueward of the Balmer edge in these runs cannot be trusted. Redward of the break, the main difference is that both full-domain runs feature absorption blueward of the Paschen edge ($\lambda=0.82\,\mu{\rm m}$). This is sensitive to the neutral population in the unconverged region, so it is likely unphysical.  

In Fig.~\ref{fig:truncated_spectra}b, we compare the H$\alpha$ profiles (analogous to Fig.~\ref{fig:halpha}). In both the full-domain and truncated runs, the lines have narrow cores and broad wings. However, in the marginally unbound wind, the absorption trough is deeper thanks to a larger column density of recombined gas. These results show that the neutral gas beyond the recombination front likely imparts important signatures onto the spectrum. This does not alter the main takeaways of this paper. 

%% FIGURE APPENDIX Comparison of Truncated Spectra

\bibliographystyle{aasjournal}
\bibliography{references}

\end{document}